\documentclass[manuscript]{aastex}

\shorttitle{Abundances in Nearby Galaxies}
\shortauthors{Pilyugin et al.}

\begin{document}


\title{The Abundance Properties of Nearby Late-Type Galaxies.\\ 
I. The Data}


\author{L.~S. Pilyugin\altaffilmark{1} and E.~K. Grebel}
\affil{Astronomisches Rechen-Institut, Zentrum f\"{u}r Astronomie 
           der Universit\"{a}t Heidelberg, 
           M\"{o}nchhofstr.\ 12--14, 69120 Heidelberg, Germany}
\email{pilyugin@mao.kiev.ua, grebel@ari.uni-heidelberg.de}

\author{A.~Y. Kniazev\altaffilmark{2,3}}
\affil{South African Astronomical Observatory, PO Box 9, 7935 Observatory, Cape Town, South Africa}
\email{akniazev@saao.ac.za}




\altaffiltext{1}{Visiting Astronomer, Main Astronomical Observatory
            of National Academy of Sciences of Ukraine,
            27 Zabolotnogo str., 03680 Kiev, Ukraine.} 
\altaffiltext{2}{Southern African Large Telescope Foundation, PO Box 9, 7935 Observatory, Cape Town, South Africa.}
\altaffiltext{3}{Sternberg Astronomical Institute, Lomonosov Moscow State University, Moscow 119992, Russia}

\begin{abstract}
We investigate the oxygen and nitrogen abundance distributions across
the optical disks of 130 nearby late-type galaxies using around 3740
published spectra of H\,{\sc ii} regions.  We use these data in order
to provide homogeneous abundance determinations for all objects in 
the sample, including H\,{\sc ii} regions in which not all of the
usual diagnostic lines were measured.  Examining the relation
between N and O abundances in these galaxies we find that the
abundances in their centres and at their isophotal $R_{25}$ disk radii
follow the same relation.  The variation in N/H at a given O/H is
around 0.3 dex.  We suggest that the observed spread in N/H may be
partly caused by the time delay between N and O enrichment and the
different star formation histories in galaxies of different
morphological types and dimensions.  We study the correlations between
the abundance properties (central O and N abundances, radial O and N
gradients) of a galaxy and its morphological type and dimension. 
\end{abstract}






\section{Introduction}

The chemical composition of a galaxy is one of its most fundamental
characteristics.  Here we focus on disk galaxies.  Because the
chemical enrichment depends on various physical processes, such as the
star formation history and the mass exchange between the galaxy and
its environment,  progress in our understanding of galaxy formation
and evolution processes depends to a large extent on improving our
knowledge of the detailed chemical properties of galaxies, such as the
radial distribution of element abundances across galactic disks.
Establishing the macroscopic properties of spiral and irregular
galaxies that likely govern the distribution of heavy elements across
their disks is very important in understanding the (chemical)
evolution of galaxies. 

Accurate abundance determinations in a sample of galaxies are
mandatory for such investigations.  The classical $T_{e}$ method,
often referred to as the direct method, is generally considered to
provide the most reliable oxygen and nitrogen abundances in H\,{\sc
ii} regions.  When this method cannot be used (due to the lack of
measurements of the required weak auroral lines in the spectra of
H\,{\sc ii} regions) then combinations of the strong nebular line
intensities in spectra of H\,{\sc ii} regions can be used as
indicators of their oxygen abundances, as was first suggested by
\citet{Pageletal1979MNRAS189} and \citet{Alloinetal1979AA78}. This
approach is usually referred to as the ``strong-line method'' and has
been widely adopted.  The establishment of calibrations (i.e., of
relations between  metallicity-sensitive emission-line combinations
and metallicity) was the subject of numerous studies \citep[][among
many
others]{DopitaEvans1986ApJ307,McGaugh1991ApJ380,Pilyugin2000AA362,
Pilyugin2001AA369,KewleyDopita2002ApJS142,
PettiniPagel2004MNRAS348,Tremonti2004ApJ613,
Liangetal2006ApJ652,Stasinska2006AA454,Thuanetal2010ApJ712}.  

A calibration is defined not only by the adopted indicators but also
by the objects that serve as calibrating data points.  A sample of
H\,{\sc ii} regions with abundances derived through the $T_e$ method
is used to construct an empirical calibration.  A set of
photoionization models is used to construct a theoretical (model)
calibration.  Even if the same indicator is used in the empirical
calibration and in the theoretical calibration and if the same spectral
measurements in a given  H\,{\sc ii} region  are used, those
calibrations can produce significantly different abundance
estimations.   Metallicities derived using theoretical calibrations
tend to be systematically higher (up to $\sim$0.7 dex) than those
derived using the empirical calibrations \citep[see reviews
by][]{KewleyEllison2008ApJ681,LopezSanchezEsteban2010AA517,LopezSanchezetal2012MNRAS426}.
Therefore, oxygen abundances in extragalactic H\,{\sc ii} regions
obtained in different studies using different calibrations can be
significantly different.

Spectroscopic measurements of H\,{\sc ii} regions within and beyond
optical radii of galaxies were carried out in many works (see list of
references below).  In these studies, usually the H\,{\sc ii} regions
in one or several galaxies are measured and the radial distributions
of the element abundances across the disks of those galaxies are
estimated.  Since often different methods for abundance determinations
are used in different works, the resulting abundances from these
studies are not homogeneous and cannot be directly compared to each
other.  Therefore, the abundances in a sample of galaxies can be
analyzed only after those abundances are homogenized, i.e., all the
abundances are redetermined in a uniform way.  This is the first step
in our present investigation.  It should be noted that  there have been 
several attempts to use uniform abundances for the determination of
radial abundance gradients in a sample of galaxies; e.g.,
\citet{VilaCostas1992MNRAS259}  for a sample of 30 galaxies,
\citet{Zaritskyetal1994ApJ420} for 39 galaxies,
\citet{vanZeeetal1998AJ116} for 11 galaxies,
\citet{Pilyuginetal2004AA425} for 54 galaxies, and
\citet{Moustakasetal2010ApJS190} for 21 galaxies.  However, those
samples contain a relatively small number of galaxies (whereas our
present sample includes 130 galaxies).

So far, little attention has been paid to the radial
distributions of nitrogen abundances in the disks of galaxies, despite
the fact that this provides  several advantages for the study of the
chemical evolution of galaxies.  Indeed, since at 12+log(O/H) $\ga$
8.3, secondary nitrogen becomes dominant and the nitrogen abundance
increases at a faster rate than the oxygen abundance
\citep[e.g.,][]{Henryetal2000ApJ541}, the change in nitrogen
abundances with galactocentric distance  should then show a larger
amplitude in comparison to oxygen abundances and, as a consequence,
should be easier to measure.  Furthermore, there is a time delay in
nitrogen production as compared to oxygen production
\citep{maeder1992,vandenhoek1997,pagel1997}.  Thus the comparison
between the radial distributions of oxygen and nitrogen abundances in
the disks of galaxies can shed additional light on the chemical
evolution of galaxies.  Therefore we consider here not only the radial
distributions of oxygen abundances but also those of nitrogen
abundances. 

Our paper is organized in the following way.  We describe the method
used for the oxygen and nitrogen abundance determinations  in the
H\,{\sc ii} regions of our galaxy sample in Section 2.  We describe
the observational data that were used to determine the abundances in
the H\,{\sc ii} regions in Section 3.  We discuss the abundance
properties in the disks of nearby galaxies (within the optical
isophotal radii) in Section 4.  We summarize our results in Section 5. 

Throughout this paper, we will use the following standard notations 
for the line intensities: \\ 
$R_2$  = $I_{\rm [O\,II] \lambda 3727+ \lambda 3729} /I_{{\rm H}\beta }$,  \\
$N_2$  = $I_{\rm [N\,II] \lambda 6548+ \lambda 6584} /I_{{\rm H}\beta }$,  \\
$S_2$  = $I_{\rm [S\,II] \lambda 6717+ \lambda 6731} /I_{{\rm H}\beta }$,  \\
$R_3$  = $I_{{\rm [O\,III]} \lambda 4959+ \lambda 5007} /I_{{\rm H}\beta }$. \\
With these definitions, the excitation parameter $P$ can be expressed as: 
$P$ = $R_3$/($R_2$+$R_3$).

\section{Abundance determination}

\subsection{Modification of the $C$ method}

A new method (called the ``$C$ method'') for oxygen and nitrogen
abundance determinations has recently been suggested
\citep{Pilyuginetal2012MNRAS424}. The idea of the $C$ method is a very
simple. We have compared a several combinations of the strong-line
intensities in the spectrum of a given H\,{\sc ii} region with those
in the spectra of a sample of reference H\,{\sc ii} regions with known
abundances in order to find the counterpart for the  H\,{\sc ii}
region under study.  It is assumed that the oxygen and nitrogen
abundances in the studied H\,{\sc ii} region are the same as in its
counterpart.  A counterpart can be selected by comparison of four
combinations of strong-line intensities:  $P$ = $R_3$/($R_2$ + $R_3$)
(excitation parameter), log$R_3$, log($N_2$/$R_2$), and
log($S_2$/$R_2$). 

However, there are recent measurements of spectra of many H\,{\sc ii}
regions where the intensities of [O\,{\sc
ii}]$\lambda$3727+$\lambda$3729 or [S\,{\sc ii}]$\lambda$6717,
[S\,{\sc ii}]$\lambda$6731) lines are not available (e.g., in
\citet{Sanchezetal2012AAa000} or in the Sloan Digital Sky Survey
(SDSS); see \citet{yorketal2000AJ120}).  It has been argued that the
oxygen and nitrogen abundances in H\,{\sc ii} regions can be estimated
even if the [S\,{\sc ii}]$\lambda$6717+$\lambda$6731 emission line is
not measured \citep{Pilyuginetal2010ApJ720} or if the [O\,{\sc
ii}]$\lambda$3727+$\lambda$3729 emission line is not available
\citep{PilyuginMattsson2011MNRAS412}.  The $C$ method can be adapted
to such cases \citep{Pilyugin2013MNRAS432}.  To find the counterpart
for the H\,{\sc ii} region under study, one does not need to compare
the four combinations of strong-line intensities, but can instead also
use only three combinations: 1) log$R_3$, $P$, and log($N_2$/$R_2$),
or 2) log$R_3$, log$N_2$, and log($N_2$/$S_2$). When this first set of
combinations of strong-line intensities is used to find the
counterpart then the resulting oxygen and nitrogen abundances will be
referred to as (O/H)$_{C_{\rm ON}}$ and (N/H)$_{C_{\rm ON}}$.  When
the second set of combinations of strong-line intensities is used to
find the counterpart then the inferred oxygen and nitrogen abundances
will be called (O/H)$_{C_{\rm NS}}$ and (N/H)$_{C_{\rm NS}}$.  

The data for reference  H\,{\sc ii} regions with $T_e$-based
abundances are compiled in \citet{Pilyuginetal2012MNRAS424}. The very
recent spectroscopic observations of
\citet{Bergetal2012ApJ754,ZuritaBresolin2012MNRAS427,Skillmanetal2013ApJ000}
have been added to the compilation. Using these combined data we
select a sample of reference H\,{\sc ii} regions for which all the
absolute differences for oxygen abundances (O/H)$_{C_{\rm ON}}$  --
(O/H)$_{T_{e}}$ and (O/H)$_{C_{\rm NS}}$  -- (O/H)$_{T_{e}}$ and for
nitrogen abundances (N/H)$_{C_{\rm ON}}$  -- (N/H)$_{T_{e}}$ and
(N/H)$_{C_{\rm NS}}$  -- (N/H)$_{T_{e}}$ are less than 0.1~dex.  This
sample of reference H\,{\sc ii} regions contains 250  objects and will
in the following be used for abundance determinations.  This sample
will be referred to as $E2013$ sample (etalon sample 2013) below.  

\subsection{Modification of the $P$ method}

Furthermore, only blue spectra were observed for H\,{\sc ii} regions
in a number of galaxies
\citep[e.g.,][]{OeyKennicutt1993ApJ411,Zaritskyetal1994ApJ420,Werketal2011ApJ735},
i.e., intensities of [N\,{\sc ii}]$\lambda$6584 and [S\,{\sc
ii}]$\lambda$6717, [S\,{\sc ii}]$\lambda$6731) are not available in
those cases.  The oxygen abundances in those H\,{\sc ii} regions can
be estimated through the $P$ calibration where only the oxygen
[O\,{\sc ii}]$\lambda$3727+$\lambda$3729 and [O\,{\sc
iii}]$\lambda$5007 lines are used
\citep{Pilyugin2000AA362,Pilyugin2001AA369,PilyuginThuan2005ApJ631}.
We have constructed a new variant of the  $P$ calibration.  The sample
of reference H\,{\sc ii} regions, $E2013$, has been used as
calibration data points. To enlarge the number of calibration data
points, we have added a number of H\,{\sc ii} regions with
(O/H)$_{C_{\rm ON}}$ abundances that  were chosen in the following
way. We determined the (O/H)$_{C_{\rm ON}}$ abundances in  H\,{\sc ii}
regions and obtained radial oxygen abundance gradients across the
disks of galaxies (see below).  The H\,{\sc ii} regions where the
deviations of the (O/H)$_{C_{\rm ON}}$ abundances from the general
radial abundance trend are less than 0.1~dex were added to the sample
of reference H\,{\sc ii} regions. 

It is well known that the relation between the oxygen abundance and
the strong oxygen line intensities is double-valued, with two distinct
parts traditionally known as the upper and lower branches of the
R$_{23}$ -- O/H diagram.   We have delimited the upper and lower
branches and the transition zone, adopting 12+log(O/H) = 8.3 as the
boundary between the upper branch and the transition zone and
12+log(O/H) = 8.0 as the boundary between the transition zone and the
lower branch. These boundaries are somewhat arbitrary, but were chosen
so as to give the best calibrations with the existing data.  Two
distinct relations between the oxygen abundance and the strong oxygen
line intensities will be established in the following, one for the
upper branch (the high-metallicity calibration) and one for   the
lower branch (the low-metallicity calibration). 

The relation between the oxygen abundance 
$Z_P$ $\equiv$ 12 + log (O/H)$_P$ and R$_{3}$ and P can be fitted by 
a polynomial of the form \citep{Pilyugin2001AA369,PilyuginThuan2005ApJ631}
\begin{equation}
Z_P = k_{0} + k_{1} \log R_3 + k_{2} (\log R_3)^2 , 
\label{equation:kR3}
\end{equation}
where we have used the notation Z $\equiv$ 12 + log (O/H) for brevity. 
To take into account the dependence on the excitation parameter $P$, 
the coefficients of Eq.~(\ref{equation:kR3}) are chosen to have  the form
\begin{equation}
k_{j} = a_{j}  + b_{j} P .
\label{equation:k}
\end{equation}
The coefficients $a_{0}$,  $a_{1}$, $a_{2}$, $b_{0}$, $b_{1}$, and 
$b_{2}$ can then be determined by looking for the best fit to our 
sample of H\,{\sc ii} regions.  We wish to derive a set of 
coefficients in Eq.~(\ref{equation:kR3}) which gives the minimum 
value of 
$\left\langle \Delta (O/H) \right\rangle$  =  
$\sqrt{(\sum\limits_{j=1}^n (\Delta (O/H)_j)^2)/n}$.
Here $\Delta (O/H)_j$ is equal to  
log(O/H)$_{P,j}$ -- log(O/H)$_{T_{e},C_{ON},j}$
for each H\,{\sc ii} region in our sample.
The quantity $\left\langle \Delta (O/H) \right\rangle$ is the average 
value of the differences between the oxygen abundances determined 
through  the $P$ calibration and the original ones. 
A few data points with large deviations, in excess of 0.15~dex, 
are rejected, and are not used in the determination of the final relation. 

The obtained upper-branch $P$ calibration (for 12 +log(O/H) $\ga$ 8.3) is 
\begin{eqnarray}
       \begin{array}{lll}
Z_P   & =  &   8.334 + 0.533\,P                                   \\
      & -  &  (0.338 + 0.415\,P)\,\log R_3                        \\
      & -  &  (0.086-0.225\,P)\,(\log R_3)^2                      \\
     \end{array}
\label{equation:ohph}
\end{eqnarray}
The obtained lower-branch $P$ calibration (for 12 +log(O/H) $\la$ 8.0)  is 
\begin{eqnarray}
       \begin{array}{lll}
Z_P   & =  &   7.949 - 1.328\,P                                  \\
      & +  & (0.926 - 0.440\,P)\,\log R_3                        \\
     & +  &  (1.220-0.480\,P)\,(\log R_3)^2                      \\
     \end{array}
\label{equation:ohpl}
\end{eqnarray}
Hence, here we use the empirical metallicity scale defined by the
H\,{\sc ii} regions with abundances derived through the direct method
($T_e$ method).

\section{The data}

We have carried out a fairly comprehensive compilation of published
spectra of H\,{\sc ii} regions in late-type galaxies.  Only those
galaxies where radial abundance gradients can be estimated were taken
into consideration.

\subsection{The general properties of our sample of galaxies}

Our final list includes 130 galaxies.  Table \ref{table:sample} lists
the general characteristics of each galaxy. The first column gives its
name. We have used the most widely used name for each galaxy. The
galaxies are listed in order of name category, with the following
categories in descending order:  \\
NGC -- New General Catalogue,       \\
IC -- Index Catalogue,     \\
UGC -- Uppsala General Catalog of Galaxies,   \\
PGC -- Catalogue of Principal Galaxies.  \\
The morphological type of the galaxy and morphological type code $T$
from {\sc leda} are reported in columns 2 and 3.  The right ascension
(R.A.) and declination (Decl.) (J2000.0) of each galaxy are given in
columns 4 and 5.  The right ascension and declination are taken from
the NASA/IPAC Extragalactic Database ({\sc ned})\footnote{The
NASA/IPAC Extragalactic Database ({\sc ned}) is operated by the Jet
Populsion Laboratory, California Institute of Technology, under
contract with the National Aeronautics and Space Administration.  {\tt
http://ned.ipac.caltech.edu/} }.  The isophotal radius $R_{25}$ in
arcmin of each galaxy is reported in column 6. Unless otherwise
stated, the $R_{25}$ values are taken from \citet[][thereafter
RC3]{RC3}.  The position angle and inclination are listed in columns 7
and 8, and the sources for these values are given in column 9.  The
adopted distance and its reference are reported in columns 10 and 11.  
The {\sc ned} distances use flow corrections for Virgo, the Great
Attractor, and Shapley Supercluster infall.  The isophotal radius in
kpc, estimated from the data in columns 6 and 10, is listed in column
12.

\subsection{A compilation of the line intensities in spectra of 
H\,{\sc ii} regions}

We have carried out an extensive search of the literature and compiled
a sample of measurements of H\,{\sc ii} regions in nearby late-type
galaxies.  We have searched for spectra of H\,{\sc ii} regions with
the requirement that they include the 
[O\,{\sc ii}]$\lambda$3727+$\lambda$3729, 
[O\,{\sc iii}]$\lambda$5007,
[N\,{\sc ii}]$\lambda$6584, and
[S\,{\sc ii}]$\lambda$6717+$\lambda$6731 lines. 
While we have tried to include as many sources as possible, we do not 
claim our search to be exhaustive. 

Thus, for each listed spectrum, we record the measured values of
[O\,{\sc ii}]$\lambda$3727+$\lambda$3729, [O\,{\sc iii}]$\lambda$5007,
[N\,{\sc ii}]$\lambda$6584, [S\,{\sc ii}]$\lambda$6717, and [S\,{\sc
ii}]$\lambda$6731.  The intensities of all lines are normalised to the
H$\beta$ line flux.  The predicted values of the flux ratio of oxygen 
[O\,{\sc iii}]$\lambda$5007/[O\,{\sc iii}]$\lambda$4959 
and nitrogen [N\,{\sc ii}]$\lambda$6584/[N\,{\sc ii}]$\lambda$6548 lines are very close
to three \citep{StoreyZeipen2000MNRAS312}. The measurements of the
[O\,{\sc iii}]$\lambda$5007 and $\lambda$4959 lines in SDSS spectra
confirm this value of the flux ratio \citep[e.g.,][]{Kniazevetal2004ApJS153}.  
Therefore, the value of $R_3$ can be estimated without  [O\,{\sc iii}]$\lambda$4959 
line as $R_3  = 1.33$[O\,{\sc iii}]$\lambda$5007, and, similarly, 
the values of $N_2$  are estimated without the lines [N\,{\sc ii}]$\lambda$6548 as 
$N_2$  = 1.33[N\,{\sc ii}]$\lambda$6584.

We have taken the de-reddened line intensities as reported by the
authors.  In some papers only the measured fluxes are reported.  In
these cases, the measured emission-line fluxes were corrected for
interstellar reddening using the theoretical H$\alpha$ to H$\beta$
ratio  (i.e., the standard  value of H$\alpha$/H$\beta$ = 2.86) and
the analytical approximation to the Whitford interstellar reddening
law from \citet{Izotovetal1994ApJ435}. 

The spectroscopic data so assembled form the basis of the present
study.  Our total list contains 3904 spectra including 162 spectra of
H\,{\sc ii} regions beyond the isophotal radius $R_{25}$.   Here we
will examine the radial oxygen and nitrogen abundance distributions
within the isophotal radius in every galaxy.  The radial oxygen and
nitrogen abundances beyond the isphotal radius will be discussed
elsewere.

\section{Abundances}

\subsection{Radial abundance gradients}

When measurements of the lines [O\,{\sc
ii}]$\lambda$3727+$\lambda$3729, [O\,{\sc iii}]$\lambda$5007, and
[N\,{\sc ii}]$\lambda$6584 were available then the oxygen
(O/H)$_{C_{\rm ON}}$ and nitrogen (N/H)$_{C_{\rm ON}}$ abundances in
the H\,{\sc ii} regions were estimated and used in our study.  If the
intensity of the line [O\,{\sc ii}]$\lambda$3727+$\lambda$3729 was not
measured but the measurements of the lines [O\,{\sc
iii}]$\lambda$5007, [N\,{\sc ii}]$\lambda$6584, [S\,{\sc
ii}]$\lambda$6717, and [S\,{\sc ii}]$\lambda$6731) were available then
the oxygen  (O/H)$_{C_{\rm NS}}$ and nitrogen (N/H)$_{C_{\rm NS}}$
abundances of the  H\,{\sc ii} regions were estimated and used.  If
the measurements of only the oxygen lines [O\,{\sc
ii}]$\lambda$3727+$\lambda$3729 and [O\,{\sc iii}]$\lambda$5007 were
available (blue spectrum was observed only) then the oxygen
(O/H)$_{P}$  abundance of the H\,{\sc ii} regions was estimated and
adopted.

The deprojected radii of the H\,{\sc ii} regions were computed using
their coordinates (or offsets from the nucleus) as reported in the
original papers, as were the position angle and inclination listed in
Table \ref{table:sample}. In some publications, the positions of the
observed H\,{\sc ii} regions (or their identifications in catalogues)
were not reported, but the deprojected radii were listed instead. In
these cases these deprojected radii were used (after correction 
for the galaxy distance adopted here, if necessary). The fractional radii
(normalized to the optical isophotal radius $R_{25}$) were obtained
with isophotal radii from Table \ref{table:sample}.  
 
The radial oxygen abundance distribution within the isophotal radius 
in every galaxy was fitted by the following equation:
\begin{equation}
12+\log({\rm O/H})  = 12+\log({\rm O/H})_{R_{0}} + C_{O/H} \times (R/R_{25}) ,
\label{equation:grado}
\end{equation} 
where 12 + log(O/H)$_{R_{0}}$ is the oxygen abundance at $R_{0}$ = 0,
i.e., the extrapolated central oxygen abundance, C$_{O/H}$, is the
slope of the oxygen abundance gradient expressed in terms of
dex~$R_{\rm 25}^{-1}$, and $R$/$R_{\rm 25}$ is the fractional radius
(the galactocentric distance normalized to the disk's isophotal radius
$R_{25}$).  We also determined the oxygen abundance gradient expressed
in terms of dex~kpc$^{-1}$.  If there were data points with large
deviations, in excess of 0.2 dex, those points were rejected, and were
not used in the determination of the final relation.

The derived parameters of the oxygen abundance distributions are
presented in Table~\ref{table:grad}.  The name of the galaxy is listed
in column 1.  The extrapolated central 12 + log(O/H)$_{R_{0}}$ oxygen
abundance and the gradient (the coefficient C$_{O/H}$ in
Eq.~(\ref{equation:grado})) expressed in terms of dex~$R_{\rm
25}^{-1}$ are listed in columns 2 and 3.  The oxygen abundance
gradient expressed in terms of dex~kpc$^{-1}$ is listed in column 4.
The scatter of oxygen abundances around the general radial oxygen
abundance trend is reported in column 5.

As in the case of the oxygen abundance, the radial nitrogen abundance 
distribution in every galaxy was fitted by the following equation:
\begin{equation}
12+\log({\rm N/H})  = 12+\log({\rm N/H})_{R_{0}} + C_{N/H} \times (R/R_{25}) . 
\label{equation:gradn}
\end{equation}
Again, if there were data points with large deviations, in excess of
0.3~dex, those points were rejected, and were not used in the
determination of the final relation.  These different rejection
criteria for oxygen abundances, 0.2~dex, and nitrogen abundances,
0.3~dex, are used for the following reason. There is no one-to-one
correspondence between nitrogen and oxygen abundances, but instead
there is a scatter in N/H at a given O/H. Therefore one can expect
that the natural scatter in N/H at a given galactocentric distance can
be larger than that in O/H assuming similar uncertainties in both
abundance determinations.  The derived parameters of the nitrogen
abundance distributions are presented in Table~\ref{table:grad}.  The
extrapolated central 12 + log(N/H)$_{R_{0}}$ nitrogen abundance and
the gradient (the coefficient C$_{N/H}$ in Eq.~(\ref{equation:gradn}))
expressed in terms of dex~$R_{\rm 25}^{-1}$ are listed in columns 6
and 7.  The nitrogen abundance gradient expressed in terms of
dex~kpc$^{-1}$ is listed in column 8.  The scatter of nitrogen
abundances around the general radial nitrogen abundance trend is
reported in column 9.

A list of references for the emission line flux measurements in the
H\,{\sc ii} regions is given in Table~\ref{table:reference}.

In the case of (O/H)$_P$ abundances, there is the following problem.
The relationship between oxygen abundance and strong oxygen line
intensities, the $P$ calibration, is double-valued with two distinct
parts usually known as the ``lower'' and ``upper'' branches of the
$R_{\rm 23}$ -- O/H relationship.  The expression for the oxygen
abundance determination in high metallicity H\,{\sc ii} regions,
Eq.~(\ref{equation:ohph}), is valid only for H\,{\sc ii} regions that
belong to the upper branch, with 12+log(O/H) $\ga$ 8.3.  Thus, one has
to know a priori on which of the two branches the H\,{\sc ii} region
lies. We can overcome this problem in the way suggested in
\citet{Pilyuginetal2004AA425}.  It has been known for a long time
\citep{Searle1971ApJ168,Smith1975ApJ199} that disks of spiral galaxies
show radial oxygen abundance gradients, in the sense that the oxygen
abundance is higher in the central part of the disk and decreases with
galactocentric distance. We thus start from the H\,{\sc ii} regions in
the central part of disks and move outward until the radius $R^*$
where the oxygen abundance decreases to 12+log(O/H) $\sim$ 8.3. It
should be noted that it is difficult to establish the exact value of
$R^*$ due to the scatter in oxygen abundance values at any fixed
radius. An unjustified use of Eq.~(\ref{equation:ohph}) in the
determination of the oxygen abundance in low-metallicity H\,{\sc ii}
regions beyond $R^*$ would result in overestimated oxygen abundances,
and would produce a false turnover in the slope of the abundance
gradients \citep{Pilyugin2003AA397}. Therefore, H\,{\sc ii} regions
with galactocentric distances larger than $R^*$, i.e., those with
12+log(O/H) $\la$ 8.3 were rejected.

The derived radial distributions of the oxygen and nitrogen abundances
in 130 galaxies are presented in  Figure~\ref{figure:gs1}. The oxygen
abundances for individual H\,{\sc ii} regions are indicated by points.
All the data points (including points with large deviations, which are
rejected and are not used in determination of the final relation) are
shown.  The linear best fits (derived via the least squares method) to
the points are represented by solid lines. The galactocentric
distances are normalized to the isophotal radius. The nitrogen
abundances for individual H\,{\sc ii} regions are shown by the plus
signs. The linear best fits to the points are indicated by dashed
lines. 

The values of the gradients in a number of galaxies (e.g., in NGC~12,
the first galaxy in our list) are rather small and are comparable (or
even lower) to the uncertainties of gradients. It can be assumed that
there is no abundance gradient in such a galaxy, and its abundance can
be specified by the  mean of the central abundance and the 
abundance at the isophotal $R_{25}$ radius of the galaxy.

Thus, the radial oxygen and nitrogen abundance distributions across
the optical disk in every galaxy are individually fitted by a single relation. It
looks like a rather good approximation for the majority of galaxies.
However, a small change in slope in the abundance distribution cannot
be excluded in the disks of several galaxies (e.g., NGC~925, NGC~3184,
NGC~5457). It is interesting to note the following.
\citet{PohlenTrujillo2006AA454} studied the surface brightness
profiles of a sample of late-type (Sb to Sdm) spiral galaxies using
imaging data from the SDSS survey. They found that the surface
brightness profiles can be divided into three classes. A small
fraction of galaxies (around 10 \%) belong to type I, which comprises
those galaxies that have a normal (standard) exponential disk down to
the noise limit.  The surface brightness distribution of the rest of
the galaxies is better described as a broken exponential.  About 60 \%
of the galaxies belong to type II, which means that they show a
down-bending profile with steeper outer part. About 30 \% of the
galaxies belong to type III, implying that they show an up-bending
profile with shallower outer part.  Thus, the abundance distribution
profile does not seem to follow strictly the slope of the surface
brightness profile.  It should be noted, however, that the number of
measured H\,{\sc ii} regions in many galaxies is too small to allow
one to detect a small bend in the radial abundance distribution.

\subsection{Notes on individual galaxies}

{\bf NGC~1512.} 
The five  H\,{\sc ii} regions that are located in the ``bridge''
between the interacting galaxies NGC~1512 and NGC~1510
\citep{Bresolinetal2012ApJ750} were excluded from our analysis.
\citet{Dicaireetal2008MNRAS385} found that there is a significant
discrepancy between the kinematically derived inclination of NGC~1512
and its photometric value ($i_{phot}$ = 65$\degr$ and $i_{kin}$ = 35$\degr$).
They noted that the explanation is quite clear: the photometric
parameter is mainly representative of the bar which contributes a
large part of the light resulting in more edge-on values. However, the
outer isophotes are much more face-on. Therefore, the kinematic value
of the inclination is adopted here.  

{\bf NGC~2442.} 
The peculiar spiral galaxy NGC~2442 is tidally distorted \citep[][and
references therein]{Pancoastetal2010ApJ723}.  

{\bf NGC~3227.} 
NGC~3227 is a nearby Seyfert galaxy that is interacting with its
gas-poor dwarf elliptical companion NGC~3226.
\citet{Mundelletal1995MNRAS277} discovered a cloud of H\,{\sc i} close
to, but physically and kinematically distinct from, the galactic disk
of NGC~3227. They suggested that this cloud (J1023+1952) might be a
dwarf galaxy that is either preexisting and being accreted by
NGC~3227, or a newly created tidal dwarf galaxy
\citep{Mundelletal1995MNRAS277,Mundelletal2004ApJ614}.  However, the
oxygen abundance of the cloud (H\,{\sc ii} regions 24 and 25 from
\citet{Werketal2011ApJ735} and of the H\,{\sc ii} region measured by
\citet{Lisenfeldetal2008ApJ685})  follows the general trend of the
radial abundance distribution in the disk of NGC~3227. The oxygen
abundances of the H\,{\sc ii} region in the cloud estimated in
different ways using the emission line measurements by
\citet{Lisenfeldetal2008ApJ685} are in satisfactory agreement to each
other: 12+log(O/H)$_{C_{ON}}$ = 8.40,  12+log(O/H)$_{C_{NS}}$ = 8.43
and  12+log(O/H)$_{P}$ = 8.34.  H\,{\sc ii} regions 10 and 12 from
\citet{Werketal2011ApJ735} follow also the general trend of the radial
abundance distribution in the disk of NGC~3227 although those H\,{\sc
ii} regions may be associated with NGC~3226. 

{\bf NGC~3239.} 
The irregular galaxy NGC~3239 is a candidate for a merging system
\citep{KrienkeHodge1990PASP102}.  The oxygen abundances determined in
four H\,{\sc ii} regions from SDSS spectra as well as the global
abundance determined using the integrated spectra from
\citet{MoustakasKennicutt2006ApJS164} are 12+log(O/H) $\sim$ 8.0
$\div$ 8.1, i.e., those abundances correspond to the transition zone
between the upper and lower branches of the O/H -- $R_{23}$ diagram.
Therefore, the (O/H)$_{P}$ abundances we derived from measurements by
\citet{Werketal2011ApJ735} are not reliable and were not used. 

{\bf NGC~3310.}
NGC~3310 is believed to be an advanced merger and its unusual outer
structure is probably the result of a recent merger with a smaller
galaxy \citep{Conseliceetal2000AJ119,KnapenJames2009ApJ698}.
Spectroscopic observations of six H\,{\sc ii} regions in NGC~3310 were
presented by \citet{Pastorizaetal1993MNRAS260} . The auroral line
[O\,{\sc iii}]$\lambda$4363 is detected in four H\,{\sc ii} regions.
The $T_e$-based abundances in those H\,{\sc ii} regions are in the
range from 12+log(O/H) = 8.13 to 8.25.  However, the coordinates of
the H\,{\sc ii} regions are not published, which prevents us from
using those H\,{\sc ii} regions in radial gradient determinations.   

{\bf NGC~3359.}
The NGC~3359 is a giant, very strongly barred spiral galaxy.
\citet{MartinRoy1995ApJ445} estimated the radial gradient using the
[O\,{\sc iii}]/H$\beta$ and  [N\,{\sc ii}]/[O\,{\sc iii}] indicators
(one-dimensional calibrations). They found the radial abundance
gradient break near the corotation radius.
\citet{ZahidBresolin2011ApJ141} carried out new measurements and
determined abundances through the $N2$ and $O3N2$ calibrations
following \citet{PettiniPagel2004MNRAS348}.  They concluded that, with
a high degree of confidence, a model with a break fits the data
significantly better than one without a break.  The $C$-based abundances
show no gradient across the entire galaxy.  The $T_e$-based abundances
in three  H\,{\sc ii} regions in NGC~3359 are in agreement with no
gradient.   

{\bf NGC~3718.} 
This galaxy is so peculiar that it is difficult to categorize it
morphologically \citep{Tullyetal1996AJ112}. 

{\bf NGC~4559.} 
The  H\,{\sc ii} regions in NGC~4559 with galactocentric radii larger
than 0.4$R_{25}$ measured by \citet{Zaritskyetal1994ApJ420} were
excluded since they are in the transition zone. 

{\bf NGC~4625.}
This is a Magellanic spiral with $R_{25}$ $\sim$ 3 kpc and an extended
faint disk reaching four times the optical $R_{25}$ radius of the
galaxy in the ultraviolet. 

{\bf NGC~5668.} 
\citet{Marinoetal2012ApJ000} have found that within $\sim36''$ 
of the nucleus the oxygen abundance O/H follows an exponential profile 
while the outer abundance trend flattens out to an approximately constant 
value and could even reverse. 

{\bf NGC 7518.} 
This is a hydrogen-deficient galaxy in the Pegasus cluster.  Following
\citet{Robertsonetal2012ApJ748}, oxygen lines from ``blue'' spectra
(360~nm -- 560~nm) have been used in the abundance determinations.  

{\bf NGC~7529.} 
This galaxy of the Pegasus cluster has a normal H\,{\sc i} content.
There is no determination of the  position angle, but the inclination
is small.  Again following \citet{Robertsonetal2012ApJ748}, oxygen
lines from ``blue'' spectra have been used in the abundance
determinations.  

{\bf NGC~7591.} 
This galaxy of the Pegasus cluster has a normal H\,{\sc i} content.
As before, following \citet{Robertsonetal2012ApJ748}, oxygen lines
from ``blue'' spectra have been used in the abundance determinations. 

{\bf IC~10.}
Due to uncertainties in the position angle and in the inclination of
IC~10, the galactocentric distances of  H\,{\sc ii} regions have been
computed without any de-projection
\citep[following][]{MagriniGongalves2009MNRAS398}. 

{\bf IC~5309.} 
A hydrogen-deficient galaxy of the Pegasus cluster.  Following
\citet{Robertsonetal2012ApJ748}, oxygen lines from ``blue'' spectra
have been used in the abundance determinations.  

{\bf UGC~9562 = II Zw 71.} 
The blue compact dwarf galaxy UGC~9562 is a probable polar-ring
galaxy, and there are a several luminous H\,{\sc ii} regions along its
major axis.  The oxygen abundances in four H\,{\sc ii} regions are
around 12+log(O/H) $\sim$ 8.2, i.e., those abundances correspond to
the transition zone between the upper and lower branches of the O/H --
$R_{23}$ diagram. Therefore, the (O/H)$_{P}$ abundances obtained with
measurements from \citet{Werketal2011ApJ735} were not used.  The
global oxygen abundances determined from the integrated spectra from
\citet{Kongetal2002AA396} are 12+log(O/H) $\sim$ 8.3.  

{\bf PGC~029167.} 
The dwarf galaxy ``Garland'' or PGC~029167 \citep[a tidal dwarf
candidate;][]{Makarovaetal2002A+A396} lies within the tidal bridges of
neutral hydrogen connecting M~81, M~82 and NGC~3077. This alignment
has led to the suggestion that this dwarf formed recently, as a result
of tidal interactions within the group. H\,{\sc ii} regions in and
near Garland exhibit enhanced metallicities compared to other galaxies
at similar luminosities.  Notably, the oxygen abundances are similar
to abundances measured in M~81 and NGC~3077 \citep[][and references
therein]{Croxalletal2009ApJ705}.

\subsection{Properties of abundances in the disks of nearby galaxies}

\subsubsection{N/H vs. O/H}

Figure~\ref{figure:ohnh} shows the nitrogen abundance as a function of
oxygen abundance in our sample of galaxies. The plus signs are the
nitrogen (N/H)$_{R_{0}}$ and oxygen abundances (O/H)$_{R_{0}}$  at
$R_{0}$ = 0, i.e., the central nitrogen and oxygen abundances. The
points show the nitrogen  (N/H)$_{R_{25}}$ and oxygen (O/H)$_{R_{25}}$
abundances at the $R_{25}$ radius of the galaxies.  Our data show a
well-defined sequence in the O/H -- N/H plot.  This sequence exhibits
the well-known turnover in the sense that the slope at low metallicity
is shallower than that at high metallicity. The commonly accepted
explanation of this change of slope is that nitrogen can be
interpreted as having both primary and secondary components.  Nitrogen
production is primary at low metallicity, but for 12 +log(O/H) $\ga$
8.3, secondary nitrogen becomes prominent, and nitrogen increases at a
faster rate than oxygen \citep{Henryetal2000ApJ541}.

Figure~\ref{figure:ohnh} shows that at high metallicity, 12 +log(O/H)
$\ga$ 8.2, the relation between the logarithms of nitrogen and oxygen
abundances can be approximated by a linear expression. For
(N/H)$_{R_{0}}$ and (O/H)$_{R_{0}}$ abundances, we found the following
relation  through the least squares method: 
\begin{eqnarray}
       \begin{array}{lll}
12+\log ({\rm N/H})_{R_{0}}  & = & 2.47\,(\pm 0.06)\,\times (12+\log ({\rm O/H})_{R_{0}})     \\
                           & - & 13.43\,(\pm  0.50) .                    \\
     \end{array}
\label{equation:nhro}
\end{eqnarray}
A similar relation 
\begin{eqnarray}
       \begin{array}{lll}
12+\log ({\rm N/H})_{R_{25}}  & = & 2.50\,(\pm 0.10)\,\times (12+\log ({\rm O/H})_{R_{25}})     \\
                           & - & 13.66\,(\pm  0.86)                     \\
     \end{array}
\label{equation:nhrg}
\end{eqnarray}
was obtained for the (N/H)$_{R_{25}}$ and (O/H)$_{R_{25}}$ abundances.  
A comparison between Eq.~(\ref{equation:nhro}) and
Eq.~(\ref{equation:nhrg}) shows that the relations N/H = $f$(O/H) for
the (N/H)$_{R_{0}}$ and (O/H)$_{R_{0}}$ abundances and for the
(N/H)$_{R_{25}}$ and (O/H)$_{R_{25}}$ abundances agree with each other
within the errors.  The  N/H = $f$(O/H) relation given by
Eq.~(\ref{equation:nhro}) is presented in  Figure~\ref{figure:ohnh} by
the solid line. The dashed lines indicate shifts along the Y-axis by
$\pm$0.15 dex.  At low metallicity, 12 +log(O/H) $\la$ 8.0, the single
relation 
\begin{equation} 12+\log ({\rm N/H}) = 0.96\,(\pm
0.08)\,\times (12+\log ({\rm O/H})) -  1.20\,(\pm  0.65)
\label{equation:nhlow} \end{equation} 
was found using both central abundances and abundances at the $R_{25}$
radii of the galaxies.  This relation is presented in
Figure~\ref{figure:ohnh} by the dotted line. 

Figure~\ref{figure:ohnh} shows that there is an appreciable spread in
N/H at a given O/H.  The variation in N/H is around a factor of 2 for
a given O/H.  The scatter can be partially attributed to the errors in
the abundance determinations but part of it seems to be true abundance
scatter.  Two major mechanisms have been proposed for generating a
spread. One mechanism invokes a time delay between ejections of the
freshly manufactured oxygen and nitrogen into the interstellar  medium
by a given stellar generation. The N/O ratios may be an indicator of
the age of a galactic system, indicating the time that has passed
since the bulk of star formation activity
\citep{EdmundsPagel1978MNRAS185}.  Thus, the N/O ratio in a galaxy
would then depend on its star formation history.  The second mechanism
for causing N/H variations at  a given O/H is a variation in the
efficiency of enriched galactic winds \citep{Pilyugin1993AA277} or/and
in the inflow of gas into the galaxy  \citep{Henryetal2000ApJ541}.  It
is believed that galactic winds do not play a significant role in the
chemical evolution of large spiral galaxies
\citep{Tremonti2004ApJ613,Dalcanton2007ApJ658}.  The enhanced N/O
ratio in individual  H\,{\sc ii} regions can be caused by the local
pollution in nitrogen by Wolf-Rayet stars \citep[][and references
therein]{LopezSanchezEsteban2010AA517}.  Since we consider ``average''
N and O abundances based on the abundances of different  H\,{\sc ii}
regions this origin of the scatter in our diagrams seems to be
unlikely.    

It has been known for a long time that galaxies of different
morphological types have different star formation histories, i.e.,
spiral galaxies with early morphological types have a larger fraction
of old stars \citep{Sandage1986AA161}. One can then expect that N/H at
a given O/H may depend on the morphological type of galaxy expressed
in terms of $T$ type \citep{Pilyuginetal2003AA397}.  On the other
hand, the star formation history of a galaxy also strongly depends on
its mass (or luminosity) as is epitomized in the galaxy downsizing
effect, where the star formation acitvity shifts from high-mass
galaxies at early cosmic times to lower-mass galaxies at later epochs
\citep{Cowieetal1996AJ112}.  Again, one can expect then that the N/H
ratio at a given O/H will depend on the galaxy mass
\citep{PilyuginThuan2011ApJ726}. 

There is a relatively tight linear correlation between the absolute
magnitudes and the logarithms of the linear diameters of nearby
galaxies \citep{vandenBergh2008AA490}.   Therefore, the linear
diameter of a galaxy can be used as indicator of its luminosity (and
mass).  Thus, if the  spread in N/H at a given O/H is caused by the
time delay between nitrogen and oxygen enrichment and the different
star formation histories in different galaxies then the N/H at a given
O/H should correlate with morphological $T$ type or/and with linear
radius of the galaxy.

Figure~\ref{figure:residual}  shows the residual of  
Eqs.~(\ref{equation:nhro}),(\ref{equation:nhrg}) as a function of the 
morphological $T$ type and linear radius of the galaxy $R_{25}$. 
Figure~\ref{figure:residual} indicates that there may exist some  
correlation between the N/H at a given O/H and morphological 
type or/and linear radius of the galaxy. To verify this, we
found a relation between N/H and O/H for spiral galaxies (with $T$
$<$ 7.5) where $T$ and log$R_{25}$ are ``secondary parameters'',  N/H
= $f$(O/H,$T$,log$R_{25}$).  It should be noted that $T$ and
log$R_{25}$ are not a perfectly independent parameters since there is
some correlation between them.  The relation obtained for central
abundances is 
\begin{eqnarray}
       \begin{array}{lll}
12+\log ({\rm N/H})_{R_{0}}  & =  &   -12.67 + 2.39\,(12+\log ({\rm O/H})_{R_{0}})     \\
      &   & -\,\,\,0.022\,T + 0.023\,\log (R_{25})  .                     \\
     \end{array}
\label{equation:nhohro}
\end{eqnarray}
The values of the coefficients in the terms containing $T$ and
log$R_{25}$ are similar. However, the $T$ value is a more important
second parameter than the log$R_{25}$ value since the variation in $T$
values (from 1 to 7.5) is much larger than the variation of
log$R_{25}$ values (from $\sim$0.6 to $\sim$1.5). Therefore, the
variation in N/H due to variation in the morphological type is larger
than that due to variation in the linear radii of galaxies.  For
example, the Sab galaxies ($T$ = 2) have nitrogen abundances larger
than
on average 0.1 dex (i.e., by around 30\%) than Sd ($T$ = 7) galaxies
with the same oxygen abundances.  The relation  N/H =
$f$(O/H,$T$,log$R_{25}$) obtained for abundances at the optical
$R_{25}$ radii of the galaxies is 
\begin{eqnarray}
       \begin{array}{lll}
12+\log ({\rm N/H})_{R_{25}}  & =  &   -14.67 + 2.59\,(12+\log ({\rm O/H})_{R_{25}})     \\
      &   & +\,\,\,0.013\,T + 0.178\,\log (R_{25})  .                       \\
     \end{array}
\label{equation:nhohrg}
\end{eqnarray}
In this case the log$R_{25}$ value is the more important second
parameter than the $T$ value.  Although the morphological type is the
more important second parameter in the relation for central abundances
while log$R_{25}$ is the more important second parameter in the
relation for abundances at the optical $R_{25}$ radii of the galaxies, 
it is difficult to draw a solid conclusion whether this is physically
meaningful since there is a correlation between $T$ and log$R_{25}$.
That the relation between N/H and O/H depends on the additional
parameter(s) $T$ and/or log$R_{25}$ suggests that the scatter  in N/H
at a given O/H can be caused, at least partly, by the time delay
between nitrogen and oxygen enrichment and the different star
formation histories in different galaxies.

\subsubsection{Abundances and gradients as a function of morphological type and galaxy radius}

The upper left panel of Figure~\ref{figure:xhrof}  shows the central
oxygen (O/H)$_{R_{0}}$ abundance in a galaxy as a function of its
morphological $T$ type (data from  Table \ref{table:sample} and
Table~\ref{table:grad}).  The (O/H)$_{R_{0}}$ -- $T$  diagram shows
that there is a trend in central oxygen abundance with morphological
type for galaxies later than Sc ($T$ $\ga$ 5) such that the central
oxygen abundances are lower in galaxies of later types.  This trend
disappears for early-type spiral galaxies ($T$ $\la$ 5). 

The upper right panel of Figure~\ref{figure:xhrof} shows the central
oxygen abundance in a galaxy as a function of its isophotal
radius $R_{25}$.  Since there is a relatively tight linear correlation
between the absolute magnitudes and the logarithms of the linear
diameters of nearby galaxies \citep{vandenBergh2008AA490}, this
diagram can be considered as some kind of analog of the standard
``luminosity -- metallicity'' diagram.  The (O/H)$_{R_{0}}$ --
$R_{25}$  diagram shows that there is a weak correlation between  the
central oxygen abundance and optical radius for small galaxies
($R_{25}$ $\la$ 10 kpc) in the sense that the smaller galaxies have on
average lower oxygen abundances.  This correlation disappears for
large galaxies ($R_{25}$ $\ga$ 10 kpc), i.e., the most oxygen-rich
galaxies of different radii have similar central oxygen abundances. 

Different versions of the luminosity -- metallicity diagram have been
constructed in earlier studies. $B$-band luminosity -- characteristic
oxygen abundance diagrams were considered in
\citet{Zaritskyetal1994ApJ420,Pilyuginetal2004AA425,Moustakasetal2010ApJS190},
where the characteristic oxygen abundance is defined as the abundance
at $R$ = 0.4$R_{25}$.  \citet{Tremonti2004ApJ613} have used the global
abundances  (in the sense that their abundances do not correspond to 
the abundances at a fixed galactocentric distance, but instead are some 
kind of mean abundance for a fraction of a galaxy within the fiber aperture) 
estimated from SDSS spectra in constructing the luminosity
-- metallicity diagram. The luminosity -- central metallicity diagram
was examined in \citet{Pilyuginetal2007MNRAS376}.  The flattening of
the luminosity -- metallicity relation at high luminosities
(essentially a plateau) can be seen in all versions of the diagram.
Thus the plateau in our (O/H)$_{R_{0}}$ -- $R_{25}$ diagram at large
radii is consistent with previous results.

It has been advocated that the constant maximum value of the observed
central oxygen abundance in the most oxygen-rich galaxies suggests
that the observed oxygen abundance in the centers of those galaxies
represents the maximum attainable value of the gas-phase oxygen
abundance \citep{Pilyuginetal2007MNRAS376}.  The upper-row panels of
Figure~\ref{figure:xhrof} show that the observed central oxygen
abundance in the most oxygen-rich galaxies in our sample is 12 +
log(O/H)$_{R_{0}}$ $\sim$ 8.85.  The dashed lines in the upper row of
panels of Figure~\ref{figure:xhrof} show this value of oxygen
abundance, which seems to correspond to the maximum attainable value
of the gas-phase oxygen abundance in galaxies.  The observed central
oxygen abundance in the most oxygen-rich galaxies from our sample is a
factor $\sim$2 higher than the gas-phase oxygen abundance in the solar
neighbourhood, 12 + log(O/H) $\sim$ 8.5
\citep[e.g.,][]{Pilyuginetal2006MNRAS367}.  The maximum attainable
value of the oxygen abundance in galaxies obtained here is in
agreement with the value from \citet{Pilyuginetal2007MNRAS376}.
Because some fraction of the oxygen (about 0.1 dex) is expected to be
locked in dust grains \citep[e.g.,][]{Estebanetal1998MNRAS295}, the
maximum value of the true (gas + dust) oxygen abundances in  H\,{\sc
ii} regions of spiral galaxies is 12 + log(O/H)$_{R_{0}}$ $\sim$ 8.95. 
 
The lower left panel of Figure~\ref{figure:xhrof} shows the central
nitrogen (N/H)$_{R_{0}}$ abundance in a galaxy disk as a function of
its morphological $T$ type.  The comparison of the panels in the left
column of Figure~\ref{figure:xhrof} shows that the changes in the
central oxygen and nitrogen abundances with morphological type of a
galaxy are rather similar.  The lower right panel of
Figure~\ref{figure:xhrof} shows the central nitrogen abundance in a
galaxy disk as a function of its optical isophotal radius $R_{25}$.
Again, the comparison between the right column panels of
Figure~\ref{figure:xhrof} indicates that the general behaviour of the
central nitrogen abundances as a function of optical radius is similar
to that for oxygen. 

According to the relation between nitrogen and oxygen abundances given
in Eq.~(\ref{equation:nhro}), the value of 12 + log(N/H)$_{R_{0}}$
$\sim$ 8.42 corresponds to the maximum attainable value of the
gas-phase oxygen abundance in galaxies with 12 + log(O/H)$_{R_{0}}$
$\sim$ 8.85.  This value of nitrogen abundance is indicated in the
lower row of panels of Figure~\ref{figure:xhrof} by the dashed lines.
One can see that the value of 12 + log(N/H)$_{R_{0}}$ $\sim$ 8.42 can
be adopted as the maximum value of the observed central nitrogen
abundance in the most nitrogen-rich galaxies at the present-day epoch.
However, in contrast to the case of oxygen, it is not necessary that
this value corresponds to the the maximum attainable value of the
gas-phase oxygen abundance in galaxies because of the time-delay
between nitrogen and oxygen enrichment of the interstellar medium. 

The panels in the left column of Figure~\ref{figure:xhrgf} show the oxygen
(O/H)$_{R_{25}}$ (upper panel) and  nitrogen (N/H)$_{R_{25}}$ (lower
panel) abundances measured at the isophotal radius, $R_{25}$, as a
function of morphological $T$ type.  A comparison between the panels
in the left column of Figure~\ref{figure:xhrof} and
Figure~\ref{figure:xhrgf} shows that the changes in the central oxygen
(nitrogen) abundances and in the abundances at the optical edge of the
disk along the Hubble sequence (or morphological type) are more or
less similar (at least qualitatively). However, the oxygen and
nitrogen abundances at the $R_{25}$ radius in four of the late-type
galaxies (NGC~4625, IC~10, UGC~223, and PGC~29167) are high and those
galaxies show a large deviation from the general trends.  NGC~4625 is
a Magellanic spiral with $R_{25}$ $\sim$ 3~kpc and a very extended
faint disk (see below).  The dwarf irregular galaxy IC~10 has a very
small optical radius, $R_{25}$ = 0.6 kpc, and a positive radial
abundance gradient. One may suggest though that any radial gradient in
such small galaxies is hard to define based on H\,{\sc ii} regions. It
remains a mystery why the oxygen abundance changes by a factor of
about two on the scale of 0.6 kpc in this galaxy.  IC~10 is one of
several dwarf galaxies in which evidence for localized, inhomogeneous
enrichment has been found
\citep{Kniazev2005AJ130,Koch2008ApJ688,Koch2008AJ135,MagriniGongalves2009MNRAS398,LopezSanchez2011MNRAS411}
-- perhaps this is a common mode of enrichment in these small objects.
In contrast, the irregular galaxy UGC~223 is rather large with
$R_{25}$ = 9.3 kpc.  The peculiar dwarf galaxy PGC~29167 (or
``Garland'') is commonly considered a tidal dwarf galaxy (see comment
on this galaxy in Section~4.2).

The panels in the right column of Figure~\ref{figure:xhrgf} show the oxygen
(O/H)$_{R_{25}}$ (upper panel) and nitrogen (N/H)$_{R_{25}}$ (lower
panel) abundances at the optical edge of the disk as defined by the
$R_{25}$ radius as a function of isophotal radius $R_{25}$.  The
positions of the galaxies in the (O/H)$_{R_{25}}$ -- $R_{25}$ and
(N/H)$_{R_{25}}$ -- $R_{25}$ diagrams do not show any obvious trends.
In particular, the (O/H)$_{R_{25}}$ and (N/H)$_{R_{25}}$ abundances do
not show any appreciable correlation with isophotal radius $R_{25}$.
However, one feature in these diagrams should be noted: the
(O/H)$_{R_{25}}$ and (N/H)$_{R_{25}}$ abundances have a maximum value in
galaxies with an isophotal radius $R_{25}$ $\sim$ 10 kpc and decrease
when moving from this value both towards smaller or larger optical
radii. 

Figure~\ref{figure:gradf} shows the radial oxygen and nitrogen
abundance gradients in units of dex~kpc$^{-1}$ as a function of
morphological $T$ type (left panels) and of isophotal radius $R_{25}$
of a galaxy (right panels).  Inspection of the left column panels of
Figure~\ref{figure:xhrgf} shows that the values of the abundance
gradients in units of dex~kpc$^{-1}$ do not correlate with the
morphological type of a galaxy.  According to
\citet{Zaritskyetal1994ApJ420}, the lack of a correlation between
gradients in units of dex~kpc$^{-1}$ and the macroscopic properties of
late-type galaxies may suggest that the relationship between these
parameters is more complex than a simple correlation. Indeed,
\citet{VilaCostas1992MNRAS259} have concluded that a correlation for
non-barred galaxies is seen.  The panels in the right column of
Figure~\ref{figure:xhrgf} show that shallow gradients can be found
both in small and large galaxies while steep gradients are seen
only in small galaxies in the sense that the smaller a galaxy the
steeper its gradient.

\section{Summary}

We compiled published spectra of H\,{\sc ii} regions in 130 nearby
galaxies. Our list contains 3904 spectra including 162 spectra  of
H\,{\sc ii} regions beyond the isophotal radius $R_{25}$.  The oxygen
and nitrogen abundances in H\,{\sc ii} regions were determined on the
metallicity scale defined by H\,{\sc ii} regions with $T_e$-based
abundances.  The radial gradients of oxygen and nitrogen abundances
across the disks of the galaxies were estimated.

At the centers of metal-rich galaxies (i.e., (12 +log(O/H) $\ga$ 8.2), we
found the relation between N and O abundances to be (N/H)$_{R_0}$
$\propto$ (O/H)$_{R_0}^{2.5}$.  The (N/H)$_{R_{25}}$  =
$f$(O/H)$_{R_{25}}$ relation between N and O abundances at the
$R_{25}$ isophotal radii of high metallicity galaxies is similar to
that for the abundances at their centers.  The variation in (N/H) at a
given (O/H) is around 0.3 dex.  To test whether the scatter in N/H
at a given O/H can be explained by the time delay between nitrogen and
oxygen enrichment and the different star formation histories in
galaxies of different morphological types and dimensions (masses), we
derived a more complex relation between N and O abundances (N/H) =
$f$((O/H),$T$,$R_{25}$).  We found that the morphological type, $T$,
is a more important ``second parameter'' in the relation for central
abundances, while the log$R_{25}$ is a more important second parameter
in the relation for abundances at the $R_{25}$ radii
of our galaxies. Since there is a correlation between $T$ and
log$R_{25}$ it is as yet unclear whether this difference is
meaningful.  The fact that the relation between N/H and O/H depends on
additional parameter(s), namely $T$ and/or log$R_{25}$ suggests that
the scatter in N/H at a given O/H can be caused, at least partly, by
the time delay between nitrogen and oxygen enrichment and the
different star formation histories in different galaxies.  The best
fit to N/H as a function of O/H is close to a linear relation at low
metallicity (12 + log(O/H) $\la$ 8.0). 

The central oxygen abundances (O/H)$_{R_0}$ show a trend along the
Hubble sequence of galaxies of late morphological types ($T$ $\ga$ 5)
such that the oxygen abundances are lower in galaxies of later types.
This trend disappears for early morphological types.  The central
oxygen abundance also correlates with optical galaxy radius for small
galaxies, $R_{25}$ $\la$ 10 kpc, being lower in galaxies of smaller
radii.  The trend disappears  for galaxies with large radii.  
The maximum gas-phase
oxygen abundance in large (10 kpc $\la$ R$_{25}$ $\la$ 30 kpc)
galaxies (or in galaxies of early (1 $\la$  $T$  $\la$ 5)
morphological types) is constant, 12 + log(O/H) $\sim$ 8.85.  This
implies that the observed central oxygen abundance of the most
oxygen-rich galaxies in our sample is a factor of $\sim$2 higher than
the gas-phase oxygen abundance in the solar neighbourhood.  The
central nitrogen abundances (N/H)$_{R_{0}}$ show a similar behaviour.
The observed central nitrogen abundance in the most nitrogen-rich
galaxies of our sample is  12 + log(O/H) $\sim$ 8.42. 

The radial O and N abundance gradients (in units of dex~kpc$^{-1}$)
within the optical radius do not show any significant correlation with
the morphological type and optical radius.  However, the spread in the
gradients increases with decreasing galaxy radius in the sense that
shallow gradients are seen both in small and large galaxies while
steep gradients occur only in a small galaxies.  The smaller a galaxy
the steeper is the gradient that it may show.

The abundance data set presented in this paper serves as the
foundation for other investigations we are carrying out. In a
forthcoming paper \citep{Pilyugin2014AJ}, we examine relations between
the radial abundance distribution across the disk and the disk
surface brightness profiles in the optical $B$ and infrared $K$ bands
for a sample of nearby galaxies.

\section*{Acknowledgements}

We are grateful to the referee for his or her constructive comments. 

L.S.P.\ and E.K.G.\ acknowledge support within the framework of
Sonderforschungsbereich (SFB 881) on ``The Milky Way System''
(especially subproject A5), which is funded by the German Research
Foundation (DFG).  L.S.P.\ thanks the Astronomisches Rechen-Institut
at Heidelberg University where this investigation was
carried out for the hospitality.  A.Y.K.\ acknowledges the support
from the National Research Foundation (NRF) of South Africa.\\ 
We thank H.J.~Zahid and F.~Bresolin for supporting us with some
unpublished details of their observations of  H\,{\sc ii} regions in
NGC~3359. \\ 

The authors acknowledge the work of the SDSS collaboration.  Funding
for SDSS-III has been provided by the Alfred P.\ Sloan Foundation, the
Participating Institutions, the National Science Foundation, and the
U.S.\ Department of Energy Office of Science. The SDSS-III web site is
http://www.sdss3.org/.  SDSS-III is managed by the Astrophysical
Research Consortium for the Participating Institutions of the SDSS-III
Collaboration including the University of Arizona, the Brazilian
Participation Group, Brookhaven National Laboratory, University of
Cambridge, Carnegie Mellon University, University of Florida, the
French Participation Group, the German Participation Group, Harvard
University, the Instituto de Astrofisica de Canarias, the Michigan
State/Notre Dame/JINA Participation Group, Johns Hopkins University,
Lawrence Berkeley National Laboratory, Max Planck Institute for
Astrophysics, Max Planck Institute for Extraterrestrial Physics, New
Mexico State University, New York University, Ohio State University,
Pennsylvania State University, University of Portsmouth, Princeton
University, the Spanish Participation Group, University of Tokyo,
University of Utah, Vanderbilt University, University of Virginia,
University of Washington, and Yale University.\\ 
We acknowledge the usage of the HyperLeda database
(http://leda.univ-lyon1.fr).



\clearpage

\setcounter{figure}{0}
\begin{figure*}
\epsscale{0.85}
\plotone{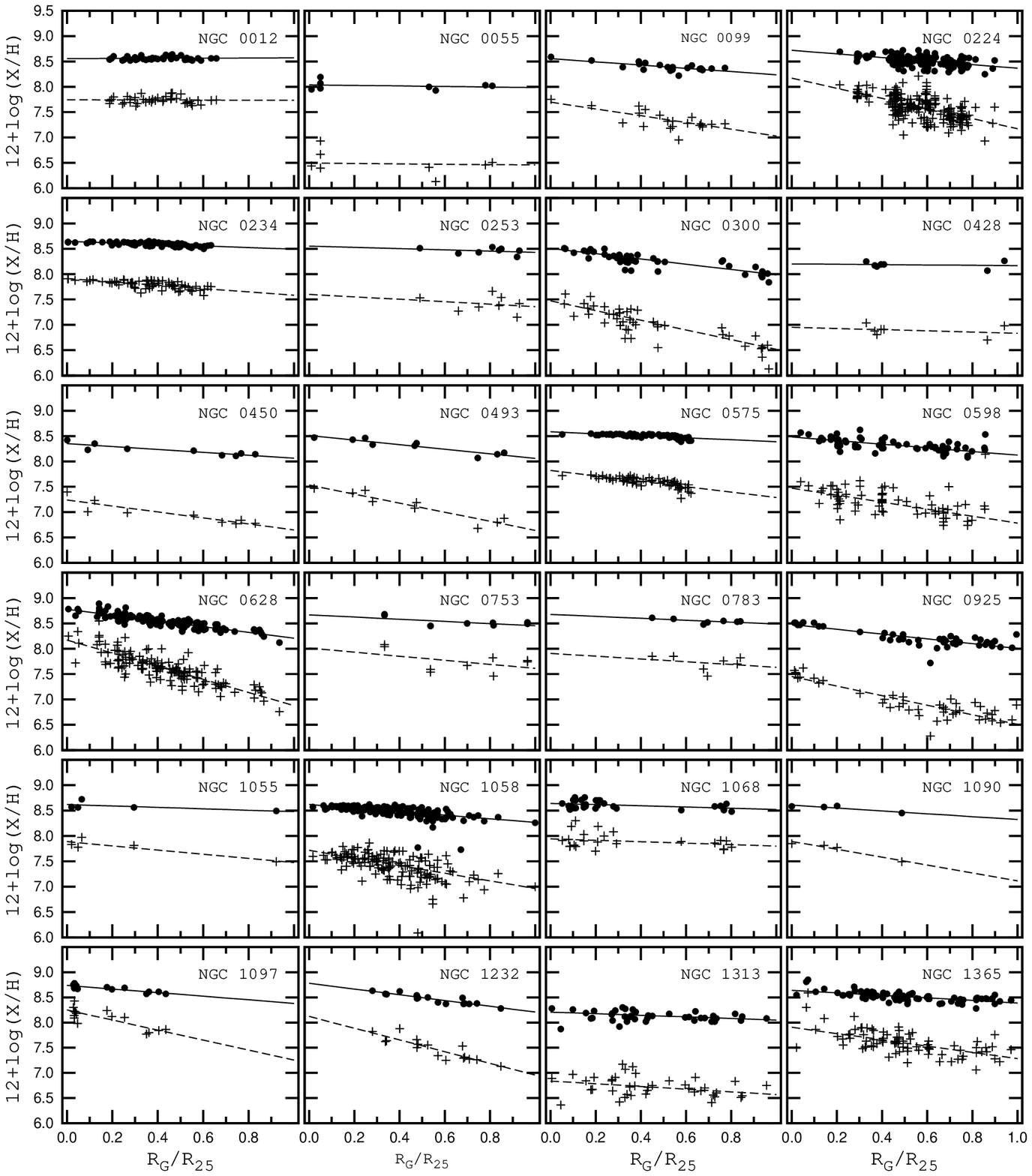}
\caption{
The radial distributions of the oxygen and nitrogen abundances 
in the disks of nearby galaxies. 
The oxygen abundances for individual H\,{\sc ii} regions 
are shown by the points. 
The linear best fits to the points 
are presented by solid lines. The galactocentric distances are normalized
to the isophotal radius.
The nitrogen abundances for individual H\,{\sc ii} regions 
are shown by the plus signs. 
The linear best fits to these  latter data (plus signs) 
are given by dashed lines. 
}
\label{figure:gs1}
\end{figure*}

\clearpage

\setcounter{figure}{0}
\begin{figure}
\epsscale{1.00}
\plotone{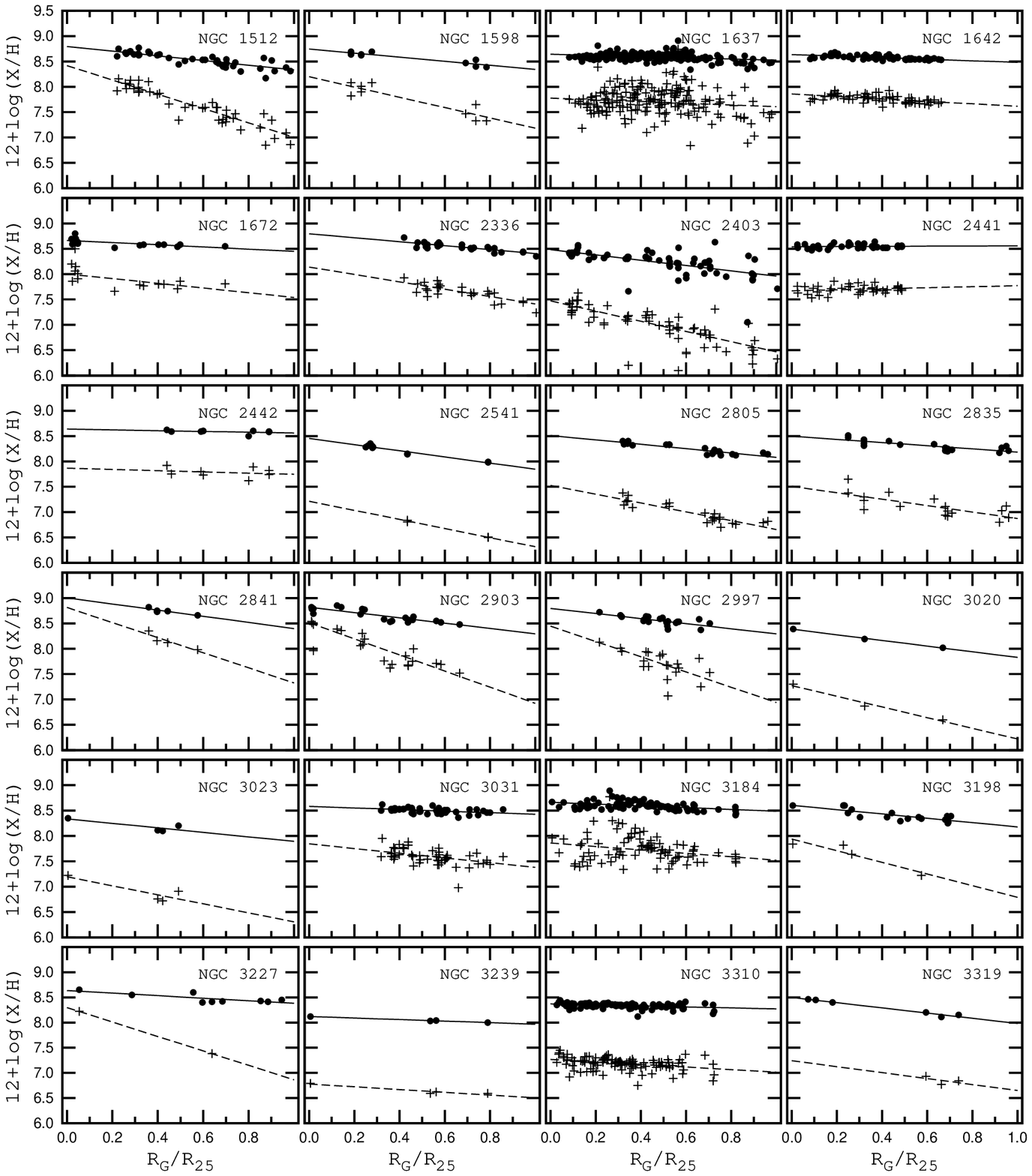}
\caption{
Continued
}
\end{figure}

\clearpage

\setcounter{figure}{0}
\begin{figure}
\epsscale{1.00}
\plotone{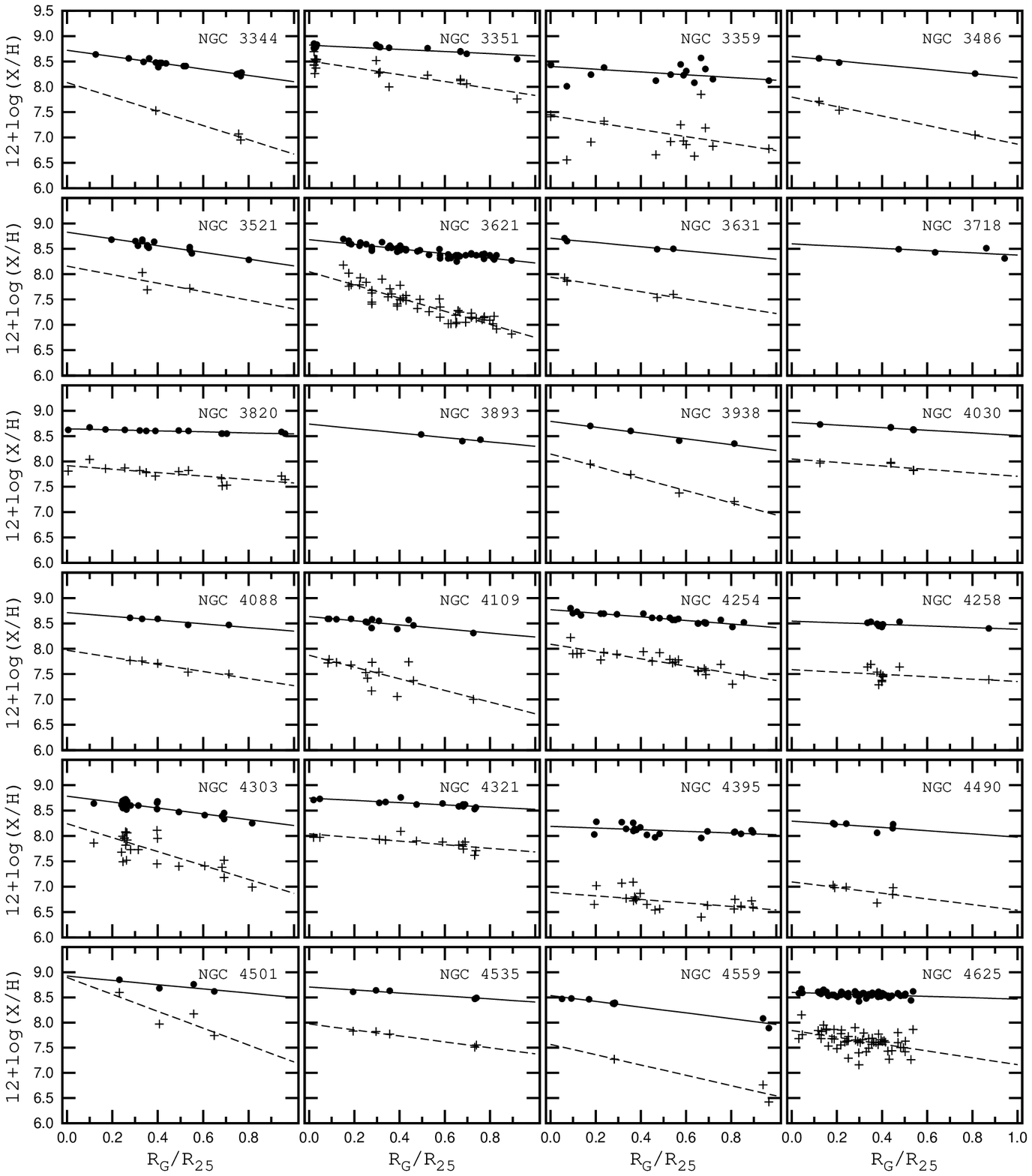}
\caption{
Continued
}
\end{figure}

\clearpage

\setcounter{figure}{0}
\begin{figure}
\epsscale{1.00}
\plotone{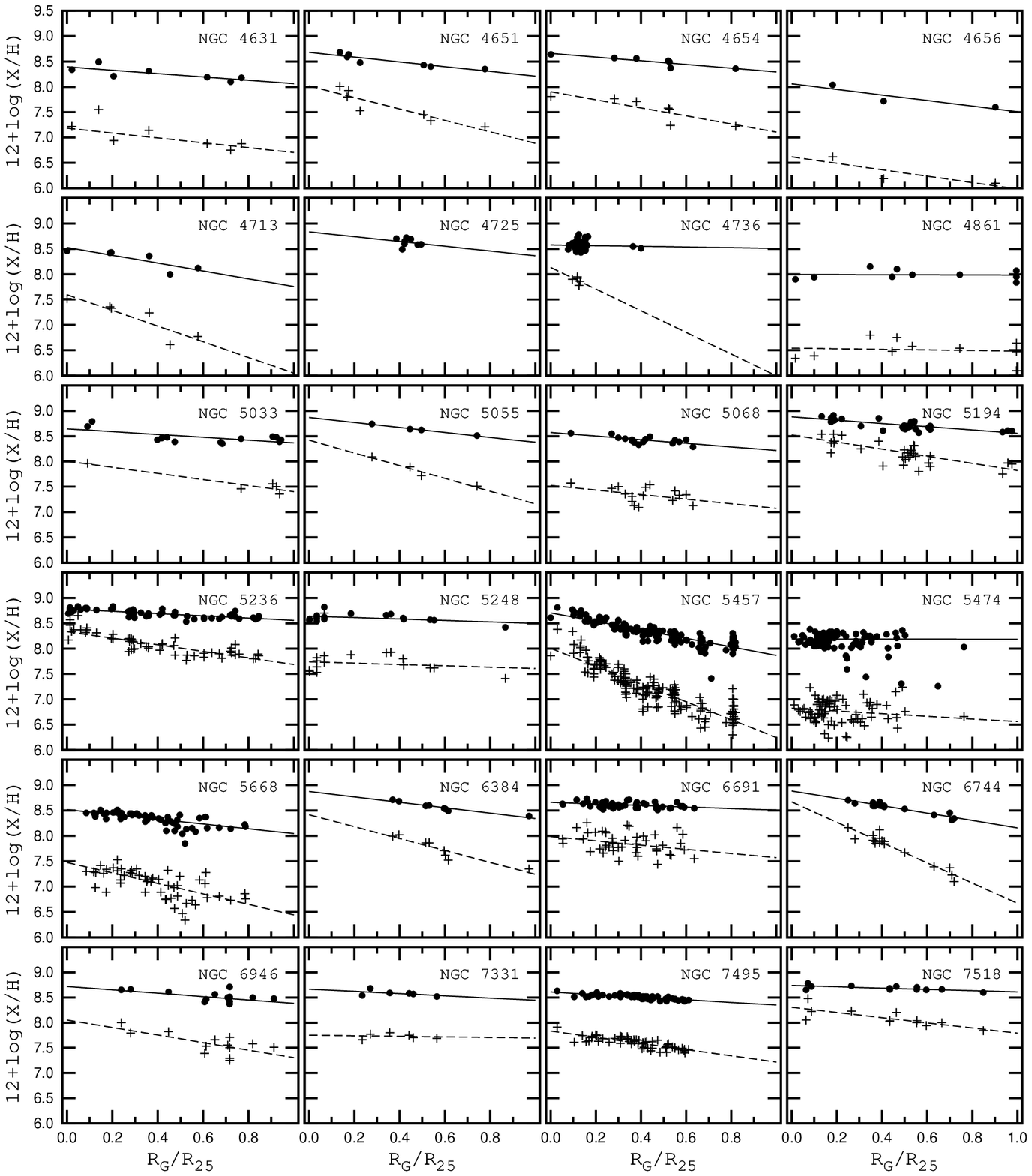}
\caption{
Continued
}
\end{figure}

\clearpage

\setcounter{figure}{0}
\begin{figure}
\epsscale{1.00}
\plotone{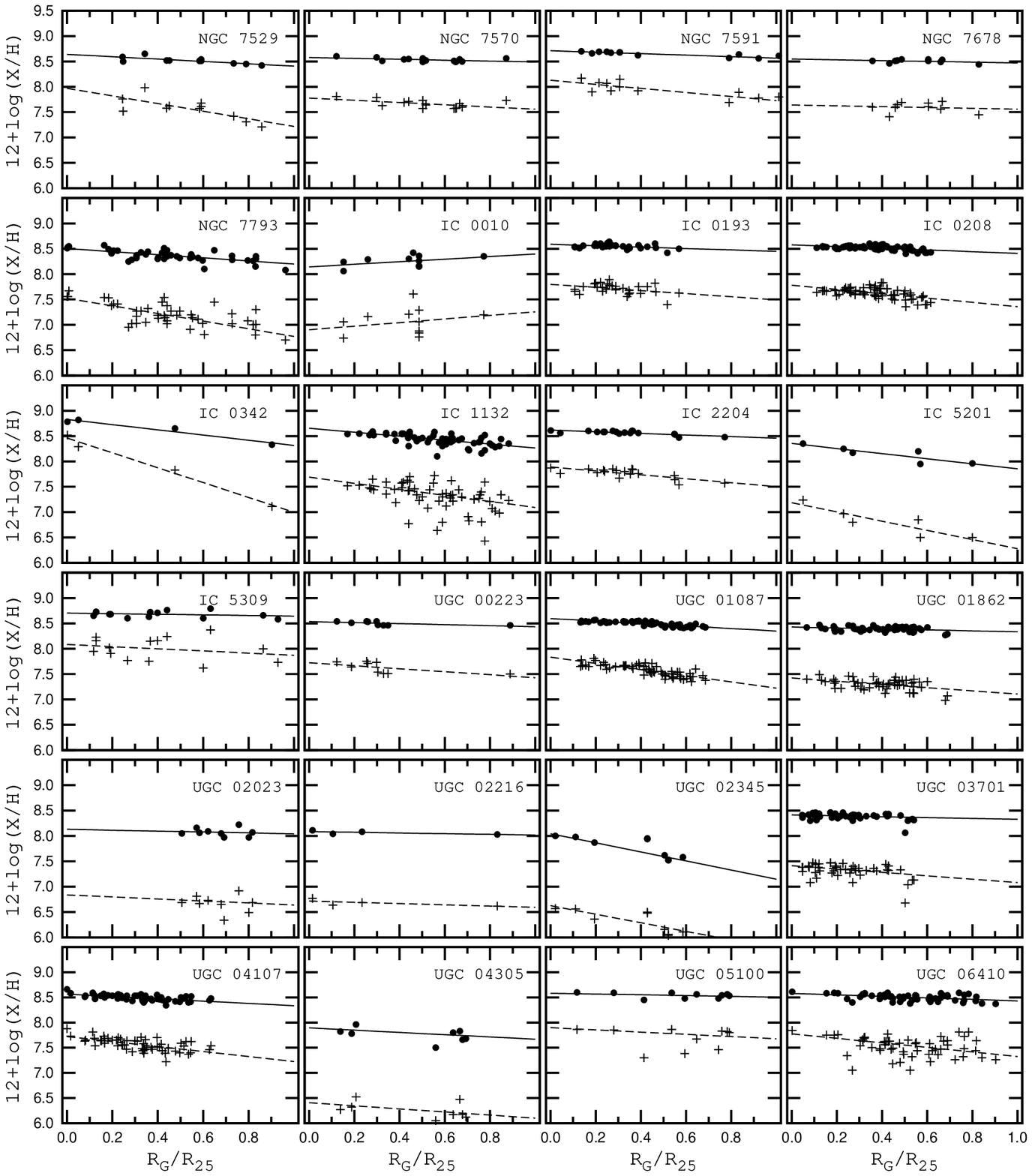}
\caption{
Continued
}
\end{figure}

\clearpage

\setcounter{figure}{0}
\begin{figure}
\epsscale{1.00}
\plotone{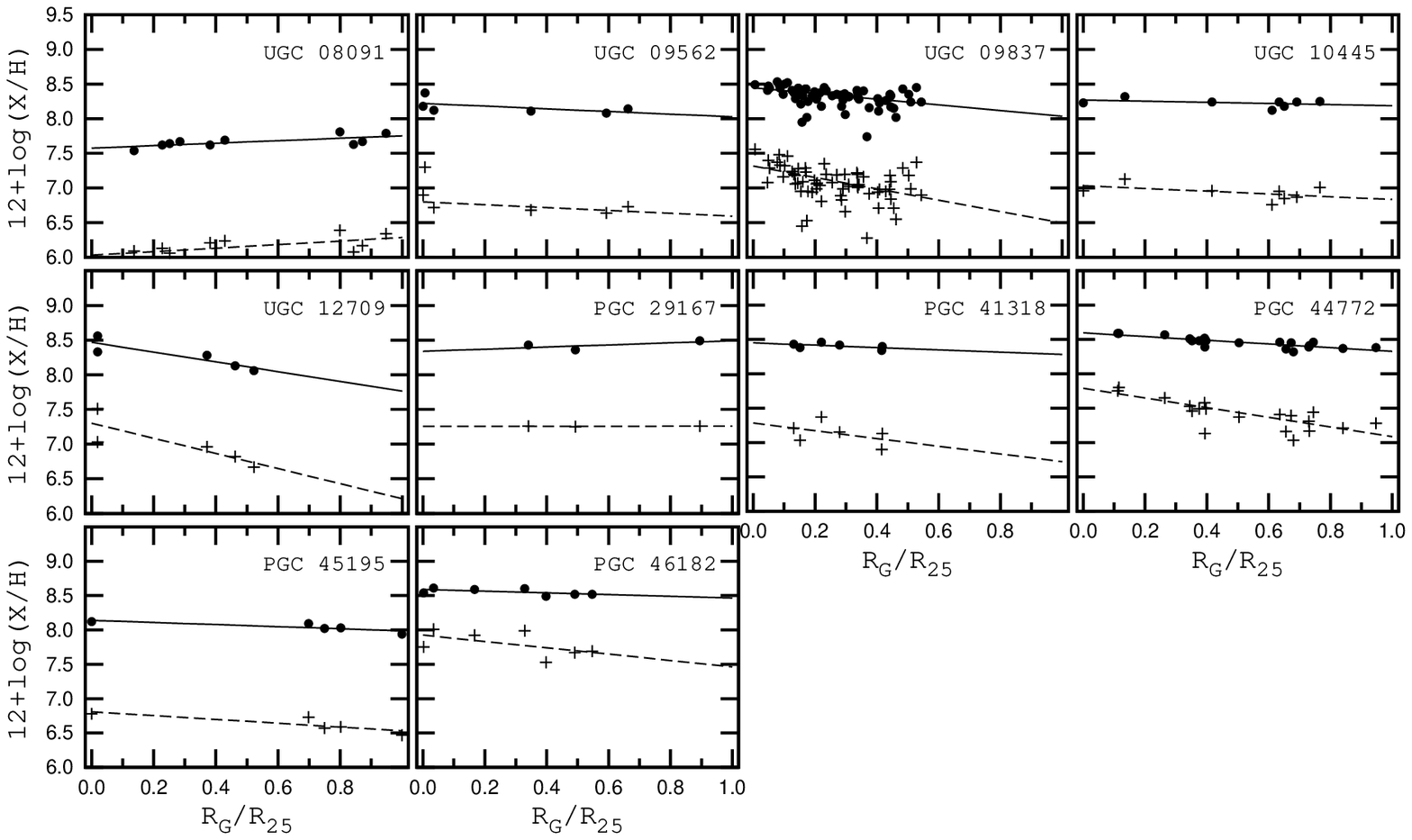}
\caption{
Continued
}
\end{figure}

\clearpage

\begin{figure}
\epsscale{1.00}
\plotone{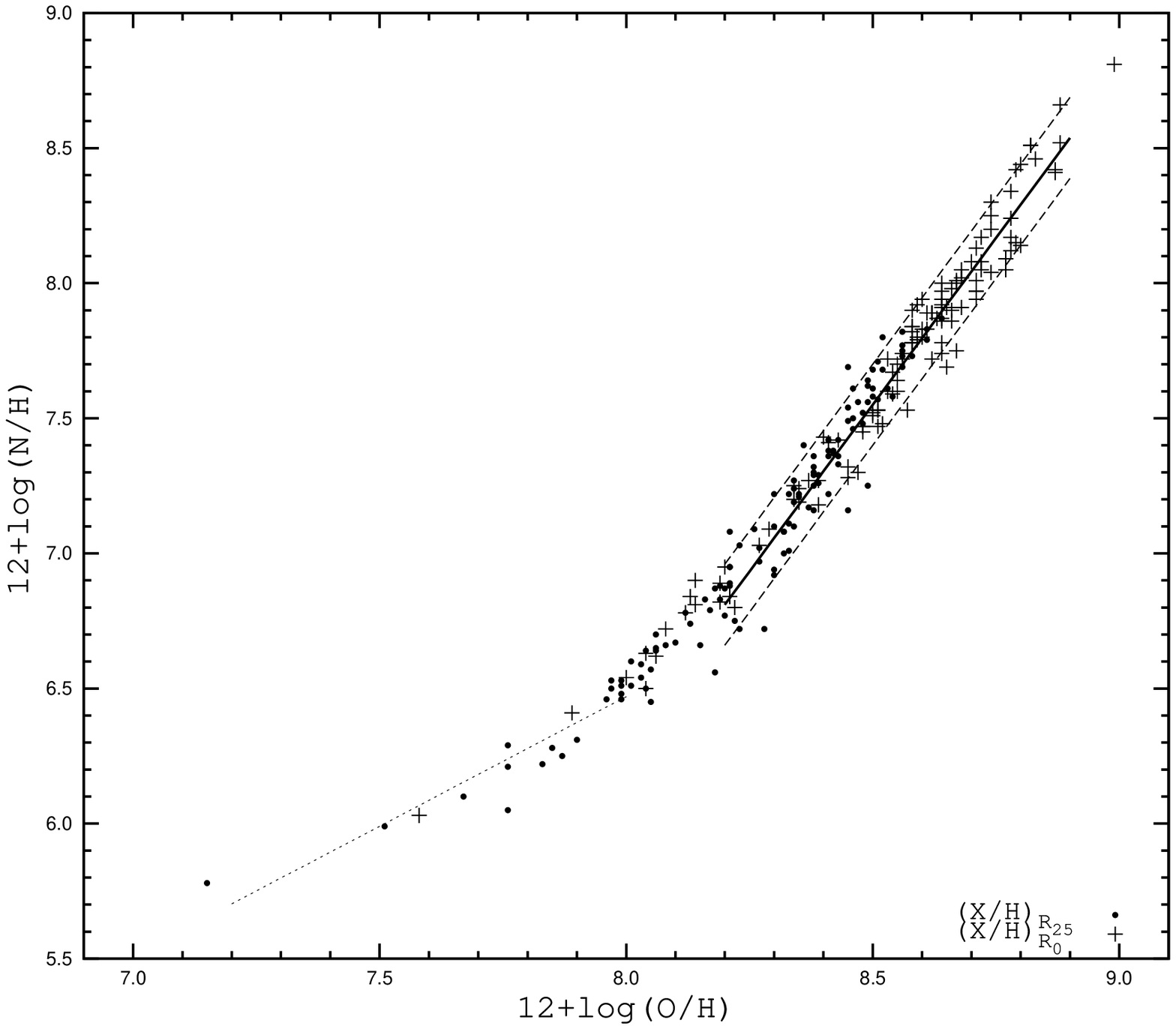}
\caption{
The N/H -- O/H diagram. 
The plus signs are the central abundances. 
The points are  abundances at the optical edges of a galaxy's $R_{25}$ isophotal radius.
The solid line is the  N/H = $f$(O/H) relation for central abundances at high 
metallicity,  Eq.~(\ref{equation:nhro}), 
the dashed lines are shifted along the Y-axis by $\pm$0.15~dex.
The dotted line is the relation at low metallicity
for both central abundances and abundances at the $R_{25}$ radii of
the galaxies. 
}
\label{figure:ohnh}
\end{figure}

\clearpage

\begin{figure*}
\epsscale{1.00}
\plotone{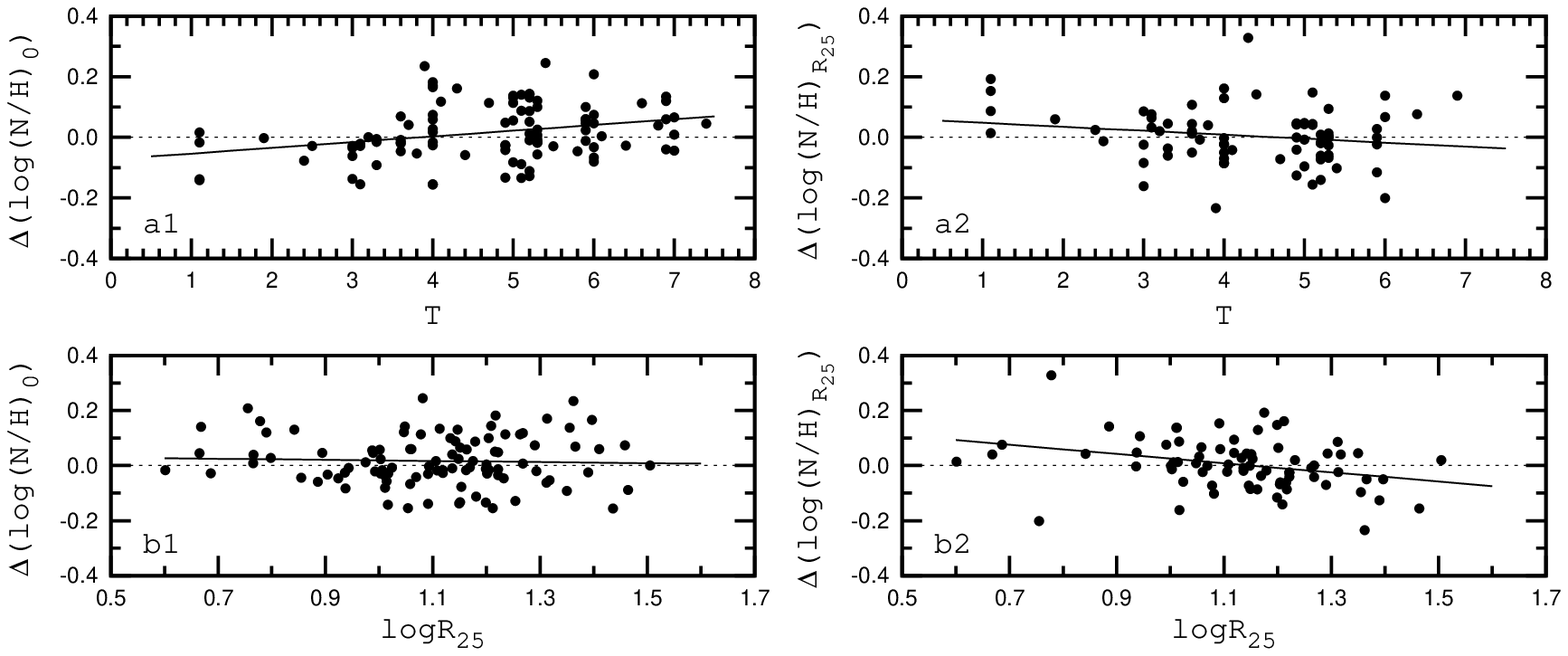}
\caption{
The residuals of Eq.~(\ref{equation:nhro}) (panels $a1$ and $b1$) and Eq.~(\ref{equation:nhrg})  
(panels $a2$ and $b2$) as a function of the morphological $T$ type and linear radius of 
the galaxy $R_{25}$. 
The points show the values of the
individual galaxies. The solid lines are linear best fits to 
those data.  The dotted lines show zero-lines.
}
\label{figure:residual}
\end{figure*}

\clearpage

\begin{figure}
\epsscale{1.00}
\plotone{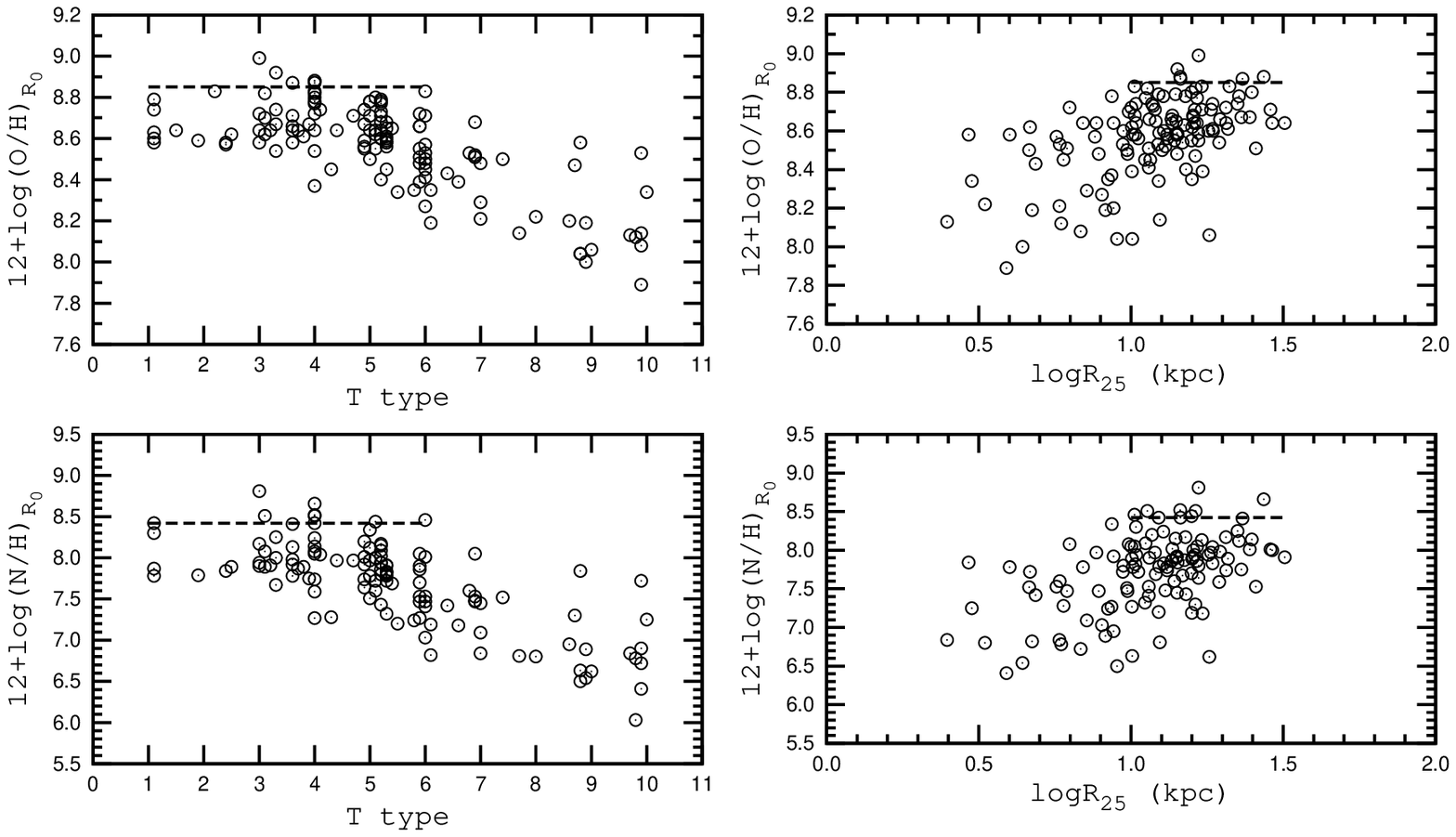}
\caption{
The central oxygen abundance, (O/H)$_{R_{0}}$, and  nitrogen abundance, (N/H)$_{R_{0}}$, 
as a function of morphological $T$ type (left panels) and of isophotal radius $R_{25}$ 
of a galaxy (right panels). The dashed lines show the maximum oxygen and nitrogen 
abundances in the galaxies.
}
\label{figure:xhrof}
\end{figure}

\clearpage

\begin{figure}
\epsscale{1.00}
\plotone{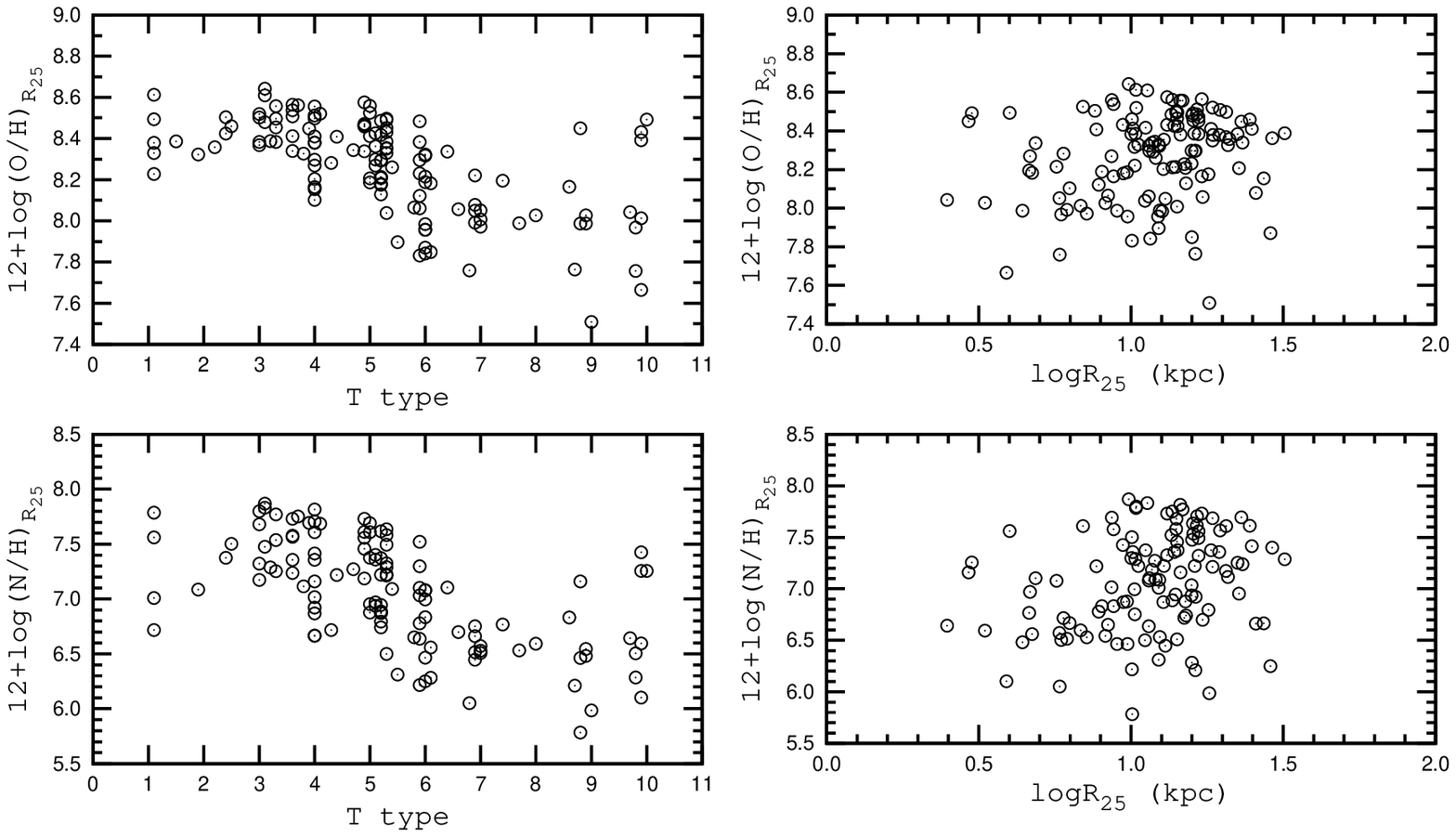}
\caption{
The oxygen abundances, (O/H)$_{R_{25}}$, and  nitrogen abundances, (N/H)$_{R_{25}}$, 
measured at the $R_{25}$ radius of the galaxies' disks 
as a function of morphological $T$ type (left panels) and of isophotal radius $R_{25}$ 
of a given galaxy (right panels). 
}
\label{figure:xhrgf}
\end{figure}

\clearpage

\begin{figure}
\epsscale{1.00}
\plotone{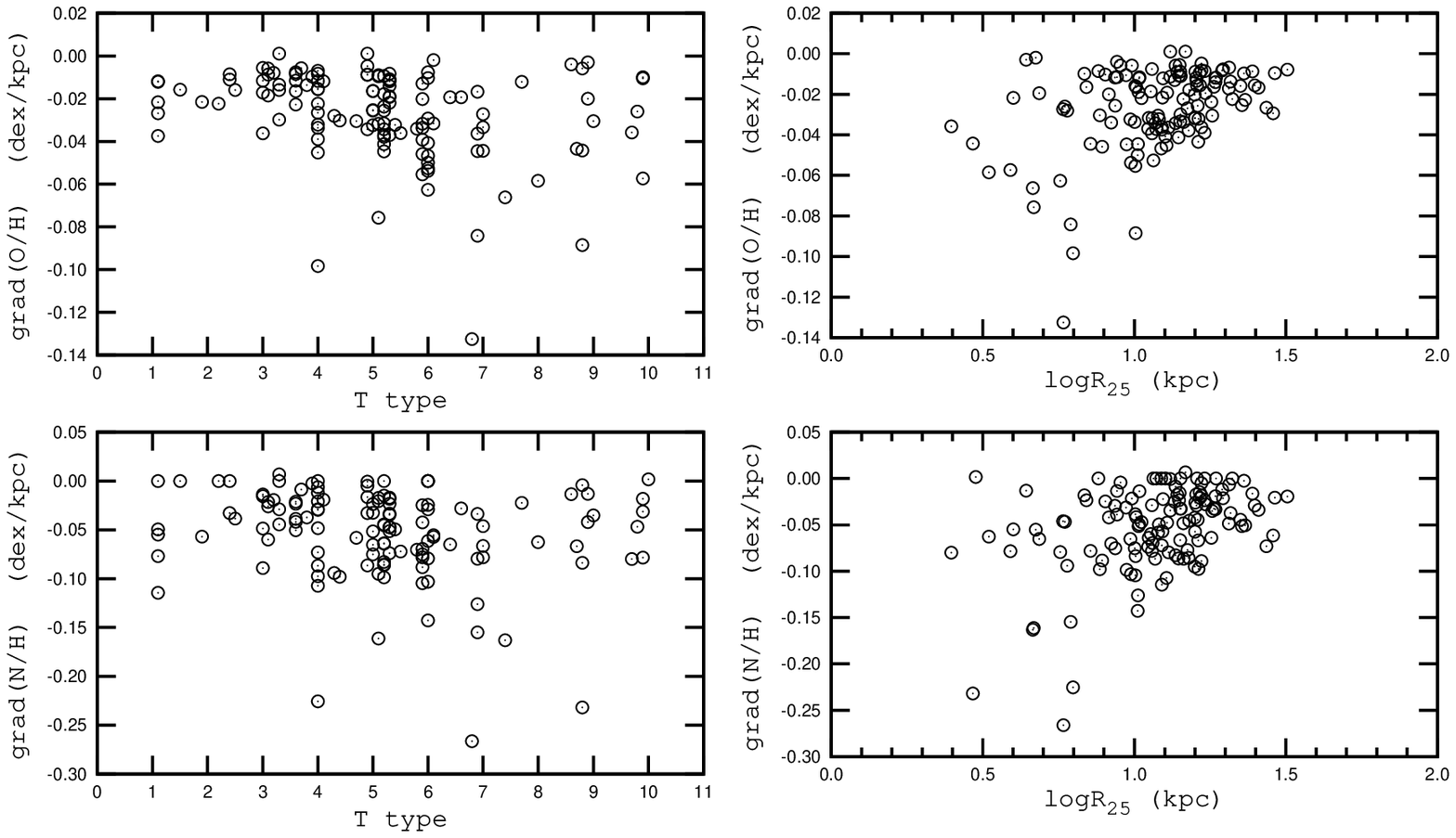}
\caption{
The radial oxygen and nitrogen abundance gradients in units of dex~kpc$^{-1}$
as a function of morphological $T$ type (left panels) and of isophotal radius $R_{25}$ 
of the galaxies (right panels).
}
\label{figure:gradf}
\end{figure}

\clearpage

\begin{deluxetable}{llcrrrrccrcr}
\tablewidth{0pt}
\tabletypesize{\scriptsize}
\rotate
\tablecaption{\label{table:sample} 
The adopted properties of our galaxies.
}
\tablehead{
\colhead{Galaxy}                &
\colhead{Type}                  &
\colhead{T}                     &
\colhead{R.A.}                  &
\colhead{Dec.}                 &
\colhead{$R_{25}$}               &
\colhead{P.A.}                  &
\colhead{Inclination}           &
\colhead{Reference}             &
\colhead{Distance}              &
\colhead{Reference}             &
\colhead{$R_{25}$}               \\ 
\colhead{}                      &
\colhead{}                      &
\colhead{}                      &
\colhead{degrees}               &
\colhead{degrees}               &
\colhead{arcmin}                &
\colhead{degrees}               &
\colhead{degrees}               &
\colhead{}                      &
\colhead{Mpc}                   &
\colhead{}                      &
\colhead{kpc}                  \\ 
}
\startdata
NGC   0012 &   Sc   &   4.9 &    2.486458 &    4.612528 &    0.85 &   125 &   27 & 1      &  53.20 & 2      &  13.14 \\ 
NGC   0055 &   SBm  &   8.8 &    3.723333 &  -39.196639 &   16.18 &   110 &   79 & 3      &   1.91 & 4      &   8.99 \\ 
NGC   0099 &   Sc   &   5.9 &    5.997470 &   15.770428 &    0.71 &     0 &   24 & 1      &  77.00 & 5      &  15.82 \\ 
NGC   0224 &   Sb   &   3.0 &   10.684792 &   41.269065 &   95.27 &    35 &   77 & 6      &   0.74 & 7      &  20.51 \\ 
NGC   0234 &   SABc &   5.3 &   10.884977 &   14.342564 &    0.81 &     0 &    0 & 1      &  59.60 & 2      &  14.06 \\ 
NGC   0253 &   SABc &   5.1 &   11.888000 &  -25.288219 &   13.77 &    50 &   76 & 8      &   3.46 & 9      &  13.86 \\ 
NGC   0300 &   Scd  &   6.9 &   13.722833 &  -37.684389 &   10.94 &   106 &   42 & 10     &   1.94 & 4      &   6.17 \\ 
NGC   0428 &   SABm &   8.6 &   18.232125 &    0.981557 &    2.04 &   120 &   41 & 1      &  14.80 & 11     &   8.77 \\
\enddata
\end{deluxetable}

\clearpage

\begin{deluxetable}{lccrrrrcc}
\tablewidth{0pt}
\tabletypesize{\scriptsize}
\rotate
\tablecaption{\label{table:grad}
The derived parameters of the radial oxygen and nitrogen abundance distributions in galaxies.  
}
\tablehead{
\colhead{Galaxy}                       &
\colhead{12+log(O/H)$_{R_{0}}$}        &
\colhead{O/H gradient}                 &
\colhead{O/H gradient}                 &
\colhead{$\sigma$(O/H)}                &
\colhead{12+log(N/H)$_{R_{0}}$}        &
\colhead{N/H gradient}                 &
\colhead{N/H gradient}                 &
\colhead{$\sigma$(N/H)}                \\
\colhead{}                             &
\colhead{}                             &
\colhead{dex~R$_{25}^{-1}$}            &
\colhead{dex~kpc$^{-1}$}               &
\colhead{dex}                          &
\colhead{}                             &
\colhead{dex~R$_{25}^{-1}$}            &
\colhead{dex~kpc$^{-1}$}               &
\colhead{dex}                          \\
}
\startdata
NGC   0012 & 8.56 $\pm$ 0.02 &  0.015 $\pm$ 0.042 &  0.0011 $\pm$ 0.0032 & 0.033 & 7.74 $\pm$ 0.04 & -0.008 $\pm$ 0.087 & -0.0006 $\pm$ 0.0066 & 0.068 \\ 
NGC   0055 & 8.04 $\pm$ 0.05 & -0.053 $\pm$ 0.094 & -0.0059 $\pm$ 0.0104 & 0.075 & 6.50 $\pm$ 0.07 & -0.037 $\pm$ 0.131 & -0.0042 $\pm$ 0.0145 & 0.091 \\ 
NGC   0099 & 8.55 $\pm$ 0.04 & -0.319 $\pm$ 0.072 & -0.0202 $\pm$ 0.0045 & 0.055 & 7.70 $\pm$ 0.07 & -0.666 $\pm$ 0.137 & -0.0422 $\pm$ 0.0087 & 0.105 \\ 
NGC   0224 & 8.72 $\pm$ 0.02 & -0.352 $\pm$ 0.039 & -0.0172 $\pm$ 0.0019 & 0.077 & 8.17 $\pm$ 0.04 & -0.996 $\pm$ 0.072 & -0.0486 $\pm$ 0.0035 & 0.136 \\ 
NGC   0234 & 8.65 $\pm$ 0.01 & -0.154 $\pm$ 0.022 & -0.0109 $\pm$ 0.0015 & 0.027 & 7.91 $\pm$ 0.02 & -0.329 $\pm$ 0.046 & -0.0234 $\pm$ 0.0033 & 0.058 \\ 
NGC   0253 & 8.55 $\pm$ 0.13 & -0.125 $\pm$ 0.164 & -0.0090 $\pm$ 0.0119 & 0.055 & 7.60 $\pm$ 0.35 & -0.241 $\pm$ 0.441 & -0.0174 $\pm$ 0.0319 & 0.149 \\ 
NGC   0300 & 8.51 $\pm$ 0.02 & -0.519 $\pm$ 0.040 & -0.0842 $\pm$ 0.0065 & 0.069 & 7.47 $\pm$ 0.04 & -0.955 $\pm$ 0.080 & -0.1548 $\pm$ 0.0129 & 0.130 \\ 
NGC   0428 & 8.20 $\pm$ 0.06 & -0.035 $\pm$ 0.109 & -0.0040 $\pm$ 0.0124 & 0.059 & 6.95 $\pm$ 0.11 & -0.118 $\pm$ 0.184 & -0.0135 $\pm$ 0.0210 & 0.099 \\
\enddata
\end{deluxetable}

\clearpage

\begin{deluxetable}{ll}
\tablewidth{0pt}
\tabletypesize{\scriptsize}
\rotate
\tablecaption{\label{table:reference}
List of references for the emission line flux measurements in the extragalactic H\,{\sc ii} regions. 
}
\tablehead{
\colhead{Galaxy}                       &
\colhead{Reference}                    \\
}
\startdata
NGC 0012          & \citet{Sanchezetal2012AAa000} \\  
NGC 0055          & \citet{WebsterSmith1983MNRAS204}, \citet{Stasinskaetal1986AA154}, \citet{Tullmannetal2003AA412}  \\  
NGC 0099          & \citet{Sanchezetal2012AAa000},  SDSS                       \\  
NGC 0224 (M~31)   & \citet{DennefeldKunth1981AJ86}, \citet{Blairetal1982ApJ254}, \citet{Galarzaetal1999AJ118}, \citet{Bresolinetal1999ApJ510}, \\
                  & \citet{Estebanetal2009ApJ700},  \citet{ZuritaBresolin2012MNRAS427}, \citet{Sandersetal2012ApJ758} \\  
NGC 0234          & \citet{Sanchezetal2012AAa000}, SDSS                       \\  
NGC 0253          & \citet{WebsterSmith1983MNRAS204} \\  
NGC 0300          & \citet{Pageletal1979MNRAS189}, \citet{WebsterSmith1983MNRAS204}, \citet{EdmundsPagel1984MNRAS211}, \citet{Bresolinetal2009ApJ700} \\
\enddata
\end{deluxetable}


\begin{thebibliography}{}

\bibitem [Alloin et al.(1979)]{Alloinetal1979AA78} 
          Alloin, D., Collin-Souffrin, S., Joly, M., \& Vigroux, L. 1979, \aap, 78, 200

\bibitem [Alloin et al.(1981)]{Alloinetal1981AA101} 
          Alloin, D., Edmunds, M. G., Lindblad, P. O., \& Pagel, B. E. J. 1981, \aap, 101, 377    

\bibitem [Amram et al.(1992)]{Amrametal1992AAS94} 
          Amram, P., Le Coarer, E., Marcelin, M., et al. 1992, \aaps, 94, 175   

\bibitem [Amram et al.(1994)]{Amrametal1994AAS103} 
          Amram, P., Marcelin, M.,  Balkowski, C., et al. 1994, \aaps, 103, 5    

\bibitem [Ball(1986)]{Ball1986ApJ307} 
          Ball, R. 1986, \apj, 307, 453   

\bibitem [Barber\'{a} et al.(2004)]{Barberaetal2004AA415} 
          Barber\'{a}, C., Athanassoula, E., \& Garc\'{i}a-G\'{o}mez C.  2004, \aap, 415, 849   

\bibitem [Barbieri et al.(2005)]{Barbierietal2005AA439} 
          Barbieri, C. V., Fraternali, F., Oosterloo, T., et al. 2005, \aap, 439, 947   

\bibitem [Begum \& Chengalur(2003)]{BegumChengalur2003AA409}   
          Begum, A., \& Chengalur, J. N. 2003, \aap, 409, 879   

\bibitem [Berg et al.(2012)]{Bergetal2012ApJ754}   
          Berg, D. A., Skillman, E. D., Marble, A. R., et al. 2012, \apj, 754, 98 

\bibitem [Bibby \& Crowther(2010)]{BibbyCrowther2010MNRAS405} 
          Bibby, J. L., \& Crowther, P. A. 2010, \mnras, 405, 2737

\bibitem [Blair et al.(1982)]{Blairetal1982ApJ254} 
          Blair, W. P., Kirshner, R. P., \& Chevalier, R. A. 1982, \apj, 254, 50 

\bibitem [Blais-Ouellette et al.(2004)]{BlaisOuelletteetal2004AA420} 
          Blais-Ouellette, S., Amram, P., Carignan, C., \& Swaters, R. 2004, \aap, 420, 147 

\bibitem [Bono et al.(2010)]{Bonoetal2010ApJ715} 
          Bono, G., Caputo, F., Marconi, M., \& Musella, I. 2010, \apj, 715, 277  

\bibitem [Boomsma et al.(2008)]{Boomsmaetal2008AA490} 
          Boomsma, R., Oosterloo, T. A., Fraternali, F., van der Hulst, J. M., \& Sancisi, R. 
          2008, \aap, 490, 555

\bibitem [Bresolin(2007)]{Bresolin2007ApJ656}
          Bresolin, F. 2007, \apj, 656, 186

\bibitem [Bresolin \& Kennicutt(2002)]{BresolinKennicutt2002ApJ572}  
          Bresolin, F., \& Kennicutt, R. C., 2002, \apj, 572, 838

\bibitem [Bresolin et al.(1999)]{Bresolinetal1999ApJ510}
          Bresolin, F., Kennicutt, R. C., \& Garnett, D. R., 1999, \apj, 510, 104

\bibitem [Bresolin et al.(2004)]{Bresolinetal2004ApJ615}
          Bresolin, F., Garnett, D. R.,  \& Kennicutt, R. C. 2004, \apj, 615, 228

\bibitem [Bresolin et al.(2005)]{Bresolinetal2005AA441}
          Bresolin, F., Schaerer, D., Gonz\'{a}lez Delgado, R. M., \& Stasi\'{n}ska, G. 2005, \aap, 441, 981

\bibitem [Bresolin et al.(2009a)]{Bresolinetal2009ApJ700}
          Bresolin, F., Gieren, W., Kudritzki, R.-P., et al. 2009a, \apj, 700, 309

\bibitem [Bresolin et al.(2009b)]{Bresolinetal2009ApJ695}  
          Bresolin, F., Ryan-Weber, E., Kennicutt, R. C., \& Goddard, Q. 2009b, \apj, 695, 580

\bibitem [Bresolin et al.(2010)]{Bresolinetal2010MNRAS404}
          Bresolin, F., Stasi\'{n}ska, G., V\'{i}lchez, J. M., Simon, J. D., \& Rosolowsky, E. 2010, \mnras, 404, 1679 

\bibitem [Bresolin et al.(2012)]{Bresolinetal2012ApJ750}
          Bresolin, F., Kennicutt, R. C., \& Ryan-Weber, E., 2012, \apj, 750, 122

\bibitem [Bush \&  Wilcots(2004)]{BushWilcots2004AJ128}
          Bush, S. J., \& Wilcots, E. M. 2004, \aj, 128, 2789  

\bibitem [Carignan(1985)]{Carignan1985ApJS58}
          Carignan, C. 1985, \apjs, 58, 107 

\bibitem [Conselice et al.(2000)]{Conseliceetal2000AJ119} 
          Conselice, C., Gallagher, J. S., Calzetti, D., Homeier, N., \& Kinney A. 2000, \aj, 119, 79

\bibitem [Cowie et al.(1996)]{Cowieetal1996AJ112} 
          Cowie, L. L., Songaila, A., Hu, E. M., \& Cohen, J. G. 1996, \aj, 112, 839 

\bibitem [Crosthwaite et al.(2000)]{Crosthwaiteetal2000AJ119}  
          Crosthwaite, L. P., Turner, J. L., \& Ho, P. T. P. 2000, \aj, 119, 1720  

\bibitem [Croxall et al.(2009)]{Croxalletal2009ApJ705} 
          Croxall, K. V., van Zee, L., Lee, H., et al. 2009, \apj, 705, 723

\bibitem [Dalcanton(2007)]{Dalcanton2007ApJ658} 
          Dalcanton, J. J. 2007, \apj, 658, 941 

\bibitem [Dalcanton et al.(2009)]{Dalcantonetal2009ApJS183}
          Dalcanton, J. J., Williams, B. F., Seth, A. C., et al. 2009, \apjs, 183, 67

\bibitem [de Blok \& Bosma(2002)]{deBlokBosma2002AA385} 
          de Blok, W. J. G., \& Bosma, A. 2002, \aap, 385, 816 

\bibitem [de Blok et al.(2008)]{deBloketal2008AJ136} 
          de Blok, W. J. G., Walter, F., Brinks, E., et al. 2008, \aj, 136, 2648 

\bibitem [Dennefeld \&  Kunth(1981)]{DennefeldKunth1981AJ86}
          Dennefeld, M., \& Kunth, D. 1981, \aj, 86, 989 

\bibitem [Dessart et al.(2008)]{Dessartetal2008ApJ675}
          Dessart, L., Blondin, S., Brown, P. J., et al. 2008, \apj, 675, 644   

\bibitem [de Vaucouleurs et al.(1991)]{RC3}
          de Vaucouleurs, G., de Vaucouleurs, A., Corvin, H. G., et al.  1991, 
          Third Reference Catalog of Bright Galaxies,  New York: Springer Verlag (RC3)

\bibitem [D\'{i}az et al.(1987)]{Diazetal1987MNRAS226} 
          D\'{i}az, A. I., Terlevich, E., Pagel, B. E. J., Vilchez, J. M., \& Edmunds, M. G. 1987, \mnras, 226, 19  

\bibitem [D\'{i}az et al.(1991)]{Diazetal1991MNRAS253} 
          D\'{i}az, A. I., Terlevich, E., Vilchez, J. M., Pagel, B. E. J., \& Edmunds, M. G. 1991, \mnras, 253, 245 

\bibitem [D\'{i}az et al.(2000)]{Diazetal2000MNRAS318} 
          D\'{i}az, A. I., Castellanos, M., Terlevich, E., \& Garc\'{i}a-Vargas, M. L. 2000, \mnras, 318, 462 

\bibitem [D\'{i}az et al.(2007)]{Diazetal2007MNRAS382} 
          D\'{i}az, \'{A}. I., Terlevich, E., Castellanos, M., \& H\"{a}gele, G. F. 2007, \mnras, 382, 251

\bibitem [Dicaire et al.(2008)]{Dicaireetal2008MNRAS385} 
          Dicaire, I., Carignan, C., Amram, P., et al. 2008, \mnras, 385, 553 

\bibitem [Dinerstein \& Shields(1986)]{DinersteinShields1986ApJ311} 
          Dinerstein, H. L., \& Shields, G. A., 1986, \apj, 311, 45

\bibitem [Dopita \& Evans(1986)]{DopitaEvans1986ApJ307}
          Dopita, M. A., \& Evans, I. N. 1986, \apj, 307, 431 

\bibitem [Drozdovsky \& Karrachentsev(2000)]{DrozdovskyKarachentsev2000AAS142} 
          Drozdovsky, I. O., \& Karachentsev, I. D. 2000, \aaps, 142, 425 

\bibitem [Dufour et al.(1980)]{Dufouretal1980ApJ236}
          Dufour, R. J., Talbot, R. J., Jensen, E. B., \& Shields, G. A. 1980, \apj, 236, 119 

\bibitem [Edmunds \& Pagel(1978)]{EdmundsPagel1978MNRAS185} 
          Edmunds, M. G., \& Pagel, B. E. J. 1978, \mnras, 185, 77 

\bibitem [Edmunds \& Pagel(1984)]{EdmundsPagel1984MNRAS211} 
          Edmunds, M. G., \& Pagel, B. E. J. 1984, \mnras, 211, 507 

\bibitem [Epinat et al.(2008)]{Epinatetal2008MNRAS390} 
          Epinat, B., Amram, P., \& Marcelin, M. 2008, \mnras, 390, 466

\bibitem [Esteban et al.(1998)]{Estebanetal1998MNRAS295}
          Esteban, C., Peimbert, M., Torres-Peimbert, S., \& Escalante, V. 1998, \mnras, 295, 401 

\bibitem [Esteban et al.(2009)]{Estebanetal2009ApJ700}
          Esteban, C., Bresolin, F., Peimbert, M., et al.  2009, \apj, 700, 654

\bibitem [Evans \& Dopita(1987)]{EvansDopita1987ApJ319} 
          Evans, I. N., \& Dopita, M. A. 1987, \apj, 319, 662      

\bibitem [Ferguson et al.(1998)]{Fergusonetal1998AJ116}  
          Ferguson, A. M. N., Gallagher, J. S., \& Wyse, R. F. G. 1998, \aj, 116, 673 

\bibitem [Fernandes et al.(2004)]{Fernandesetal2004MNRAS355}  
          Fernandes, I. F., de Carvalho, R., Contini, T., \& Gal, R. R. 2004, \mnras, 355, 728

\bibitem [Fierro et al.(1986)]{Fierroetal1986PASP98} 
          Fierro, J., Torres-Peimbert, S., \& Peimbert, M. 1986, \pasp, 98, 1032

\bibitem [Firpo et al.(2005)]{Firpoetal2005MNRAS356} 
          Firpo, V., Bosch, G., \& Morrell, N. 2005, \mnras, 356, 1357 

\bibitem [Galarza et al.(1999)]{Galarzaetal1999AJ118}  
          Galarza, V. C., Walterbos, R. A. M., \& Braun, R. 1999, \aj, 118, 2775 

\bibitem [Garc\'{i}a-Benito et al.(2010)]{GarciaBenitoetal2010MNRAS408}
          Garc\'{i}a-Benito, R., D\'{i}az, A. I.,  H\"{a}gele, G. E., et al. 2010, \mnras, 408, 2234 

\bibitem [Garc\'{i}a-G\'{o}mez et al.(2004)]{GarciaGomezetal2004AA421}  
          Garc\'{i}a-G\'{o}mez, C., Barber\'{a}, C., Athanassoula, E., Bosma, A., \& Whyte, L. 
          2004, \aap, 421, 595  

\bibitem [Garnett \& Kennicutt(1994)]{GarnettKennicutt1994ApJ426123}  
          Garnett, D. R., \& Kennicutt, R. C. 1994, \apj, 426, 123

\bibitem [Garnett et al.(2004)]{Garnettetal2004ApJ607}  
           Garnett, D. R.,  Kennicutt, R. C., \& Bresolin, F. 2004, \apj, 607, L21

\bibitem [Garnett \& Shields(1987)]{GarnettShields1987ApJ317}  
          Garnett, D. R., \& Shields, G. A. 1987, \apj, 317, 82 

\bibitem [Garnett et al.(1999)]{Garnettetal1999ApJ489}  
          Garnett, D. R., Shields, G. A., Peimbert, M., et al. 1999, \apj, 513, 168

\bibitem [Garnett et al.(1997)]{Garnettetal1997ApJ489}  
          Garnett, D. R., Shields, G. A., Skillman, E. D., Sagan, S. P., \& Dufour, R. J. 
          1997, \apj, 489, 63
 
\bibitem [Gil de Paz et al.(2007)]{GildePazetal2007ApJS173} 
           Gil de Paz, A., Boissier, S., Madore, B. F., et al. 2007, \apjs, 173, 185 
 
\bibitem [Goddard et al.(2011)]{Goddardetal2011MNRAS412}  
          Goddard, Q., Bresolin, F., Kennicutt, R. C., Ryan-Weber, E. V., \& Rosales-Ortega, F. F. 
          2011, \mnras, 412, 1246

\bibitem [Gonz\'{a}lez Delgado \& P\'{e}rez(1997)]{GonzalezDelgado1997MNRAS284} 
          Gonz\'{a}lez Delgado, R. M., \& P\'{e}rez, E. 1997, \mnras, 284, 931 

\bibitem [Grosb\o l(1985)]{Grosbol1985AAS60}  
          Grosb\o l, P. J.  1985, \aaps, 60, 261

\bibitem [Gusev et al.(2012)]{Gusevetal2012MNRAS424}  
          Gusev, A. S., Pilyugin, L. S., Sakhibov, F., et al. 2012, \mnras, 424, 1930

\bibitem [Hadfield \& Crowther(2007)]{HadfieldCrowther2007MNRAS381}  
          Hadfield, L. J., \& Crowther, P. A. 2007, \mnras, 381, 418  

\bibitem [Hawley (1978)]{Hawley1978ApJ224}  
          Hawley, S. A. 1978, \apj, 224, 417  

\bibitem [Henry et al.(1994)]{Henryetal1994MNRAS266}  
          Henry, R. B. C., Pagel, B. E. J., \& Chincarini, G. L. 
          1994, \mnras, 266, 421

\bibitem [Henry et al.(1996)]{Henryetal1996MNRAS283}  
          Henry, R. B. C., Balkowski, C., Cayatte, V., Edmunds, M. G., \& Pagel, B. E. J. 
          1996, \mnras, 283, 635

\bibitem [Henry et al.(2000)]{Henryetal2000ApJ541} 
          Henry, R. B. C., Edmunds, M. G., \& K\'{o}ppen, J. 2000, \apj, 541, 660

\bibitem [Herrmann \& Ciardullo(2009)]{HerrmannCiardullo2009ApJ705}  
          Herrmann, K. A., \& Ciardullo, R. 2009, \apj, 705, 1686 

\bibitem [Herrmann et al.(2008)]{Herrmannetal2008ApJ683}  
          Herrmann, K. A., Ciardullo, R., Feldmeier, J. J., \& Vinciguerra, M. 2008, \apj, 683, 630  

\bibitem [Hess et al.(2009)]{Hessetal2009ApJ699}  
          Hess, K. M., Pisano, D. J., Wilcots, E. M., \& Chengalur, J. N. 2009, \apj, 699, 76 

\bibitem [Hlavacek-Larrondo et al.(2011)]{HlavacekLarrondoetal2011MNRAS411}  
          Hlavacek-Larrondo, J., Carignan, C., Daigle, O., et al. 2011, \mnras, 411, 71

\bibitem [Humphreys et al.(2008)]{Humphreysetal2008ApJ672}  
          Humphreys, E. M. L., Reid, M. J., Greenhill, L. J., Moran, J. M., \& Argon, A. L. 
          2008, \apj, 672, 800  

\bibitem [Izotov et al.(1994)]{Izotovetal1994ApJ435}
          Izotov, Y. I., Thuan, T. X., \& Lipovetsky, V. A. 1994, \apj, 435, 647

\bibitem [Izotov et al.(1997)]{Izotovetal1997ApJS108}
          Izotov, Y. I., Thuan, T. X., \& Lipovetsky, V. A. 1997, \apjs, 108, 1  

\bibitem [Izotov et al.(2007)]{Izotovetal2007ApJ662}
          Izotov, Y. I., Thuan, T. X., \& Stasi\'{n}ska, G. 2007, \apj, 662, 15 

\bibitem [Jacobs et al.(2009)]{Jacobsetal2009AJ138}
          Jacobs, B. A., Rizzi, L., Tully, R. B., et al. 2009, \aj, 138, 332  

\bibitem [Jamet et al.(2005)]{Jametetal2005AA444}
          Jamet, L., Stasi\'{n}ska, G., P\'{e}rez, E., Gonz\'{a}lez Delgado, R. M., \& V\'{i}lchez, J. M.
          2005, \aap, 444, 723

\bibitem [Kamphuis(1993)]{Kamphuis1993PhD} 
          Kamphuis, J. J. 1993, PhD Thesis, Univ. Groningen 

\bibitem [Kamphuis \& Briggs(1992)]{KamphuisBriggs1992AA253} 
          Kamphuis, J., \& Briggs, F. 1992, \aap, 253, 335       

\bibitem [Karachentsev et al.(2002)]{Karachentsevetal2002AA383} 
          Karachentsev, I. D., Dolphin, A. E., Geisler, D., et al. 2002, \aap, 383, 125 

\bibitem [Kehrig et al.(2011)]{Kehrigetal2011AA526} 
          Kehrig, C., Ory, M. S., Crowther, P.A., et al. 2011, \aap, 526, A128 

\bibitem [Kennicutt et al.(2003)]{Kennicuttetal2003ApJ591}
          Kennicutt, R. C., Bresolin, F., \& Garnett, D. R. 2003, \apj, 591, 801

\bibitem [Kennicutt \& Garnett(1996)]{KennicuttGarnett1996ApJ456} 
          Kennicutt, R. C., \& Garnett, D. R. 1996, \apj, 456, 504 

\bibitem [Kewley \& Dopita(2002)]{KewleyDopita2002ApJS142}
          Kewley, L. J., \& Dopita, M. A. 2002, \apjs, 142, 35

\bibitem [Kewley \& Ellison(2008)]{KewleyEllison2008ApJ681} 
          Kewley, L. J., \& Ellison, S. L. 2008, \apj, 681, 1183

\bibitem [Kinkel \& Rosa(1994)]{KinkelRosa1994AA282} 
          Kinkel, U., \& Rosa, M. R. 1994, \aap, 282, L37

\bibitem [Kirby et al.(2008)]{Kirbyetal2008AJ136} 
          Kirby, E. M., Jerjen, H., Ryder, S. D., \& Driver, S. P. 2008, \aj, 136, 1866  

\bibitem [Knapen \& James(2009)]{KnapenJames2009ApJ698}  
          Knapen, J. H., \& James, P. A. 2009, \apj, 698, 1437

\bibitem [Kniazev et al.(2004)]{Kniazevetal2004ApJS153}  
          Kniazev, A. Y., Pustilnik, S. A., Grebel, E. K., Lee, H., \& Pramskij, A. G. 
          2004, \apjs, 153, 429

\bibitem [Kniazev et al.(2005)]{Kniazev2005AJ130} 
          Kniazev, A. Y., Grebel, E. K., Pustilnik, S. A., Pramskij, A. G., \& Zucker, D. B.  
          2005, \aj, 130, 1558

\bibitem [Kobulnicky \& Skillman(1998)]{KobulnickySkillman1998ApJ497} 
          Kobulnicky, H. A., \& Skillman, E. D. 1998, \apj, 497, 601 

\bibitem [Koch et al.(2008a)]{Koch2008AJ135} 
          Koch, A., Grebel, E. K., Gilmore, G. F., et al. 2008a, \aj, 135, 1580

\bibitem [Koch et al.(2008b)]{Koch2008ApJ688} 
          Koch, A., McWilliam, A.,  Grebel, E. K., Zucker, D. B., \& Belokurov, V. 
          2008b, \apj, 688, L13

\bibitem [Kong et al.(2002)]{Kongetal2002AA396} 
          Kong, X., Cheng, F. Z., Weiss, A., \& Charlot, S. 2002, \aap, 396, 503

\bibitem [Kopparapu et al.(2008)]{Kopparapuetal2008ApJ675} 
          Kopparapu, R. K., Hanna, C., Kalogera, V., et al. 2008, \apj, 675, 1459                                            

\bibitem [Krienke \& Hodge(1990)]{KrienkeHodge1990PASP102}  
          Krienke, K., \& Hodge, P. 1990, \pasp, 102, 41 

\bibitem [Kwitter \& Aller(1981)]{KwitterAller1981MNRAS195} 
          Kwitter, K. B., \& Aller, L. H. 1981, \mnras, 195, 939  

\bibitem [Lee et al.(2011)]{Leeetal2011ApJS192} 
          Lee, J. C., Gil de Paz, A., Kennicutt, R. C., et al. 2011, \apjs, 192, 6 

\bibitem [Lequeux et al.(1979)]{Lequeuxetal1979AA80} 
          Lequeux, J., Peimbert, M., Rayo, J. F., Serrano, A., \& Torres-Peimbert, S. 
          1979, \aap, 80, 155  

\bibitem [Liang et al.(2006)]{Liangetal2006ApJ652}  
          Liang, Y. C., Yin, S. Y., Hammer, F., et al. 2006, \apj, 652, 257

\bibitem [Lisenfeld et al.(2008)]{Lisenfeldetal2008ApJ685} 
          Lisenfeld, U., Mundell, C. G., Schinnerer, E., Appleton, P. N., \& Allsopp, J. 
          2008, \apj, 685, 181  

\bibitem [L\'opez-S\'anchez \& Esteban (2010)]{LopezSanchezEsteban2010AA517}
          L\'opez-S\'anchez, \'A. R., \& Esteban, C. 2010, \aap, 517, A85

\bibitem [L{\'o}pez-S{\'a}nchez et al.(2011)]{LopezSanchez2011MNRAS411} 
          L{\'o}pez-S{\'a}nchez, {\'A}. R., Mesa-Delgado, A., 
          L{\'o}pez-Mart{\'{\i}}n,  L., \& Esteban, C. 2011, \mnras, 411, 2076 

\bibitem [L\'opez-S\'anchez et al.(2012)]{LopezSanchezetal2012MNRAS426}
          L\'opez-S\'anchez, \'A. R., Dopita, M. A., Kewley, L. J., et al. 2012, \mnras, 426, 2630

\bibitem [Luridiana et al.(2002)]{Luridianaetal2002RevMex38}
          Luridiana, V., Esteban, C., Peimbert, M., \& Peimbert, A. 2002,
          Rev. Mex. A. A., 38, 97 

\bibitem [Maeder(1992)]{maeder1992}
          Maeder, A. 1992, \aap, 264, 105

\bibitem [Magrini \& Gon{\c c}alves(2009)]{MagriniGongalves2009MNRAS398}  
          Magrini, L., \& Gon{\c c}alves, D. R. 2009, \mnras, 398, 280 

\bibitem [Magrini et al.(2010)]{Magrinietal2010AAA63}
          Magrini, L., Stanghellini, L., Corbelli, E., Galli, D., \& Villaver, E.
          2010, \aap, 512, A63 

\bibitem [Makarova et al.(2002)]{Makarovaetal2002A+A396} 
          Makarova, L. N., Grebel, E. K., Karachentsev, I. D., et al. 2002, \aap, 396, 473

\bibitem [Marcelin \& Athanassoula(1982)]{MarcelinAthanassoula1982AA105}
          Marcelin, M., \& Athanassoula, E. 1982, \aap, 105, 76  

\bibitem [Marcelin et al.(1983)]{Marcelinetal1983AA128}
          Marcelin, M., Boulesteix, J., \& Georgelin, Y. 1983, \aap, 128, 140 

\bibitem [Marino et al.(2012)]{Marinoetal2012ApJ000} 
          Marino, R. A., Gil de Paz, A., Castillo-Morales, A., et al. 2012, \apj, 754, 61 

\bibitem [Martin \& Roy(1995)]{MartinRoy1995ApJ445} 
          Martin, P., \& Roy, J.-R. 1995, \apj, 445, 161 

\bibitem [McCall et al.(1985)]{McCalletal1985ApJS57}
          McCall, M. L., Rybski, P. M., \& Shields, G. A. 1985, \apjs, 57, 1

\bibitem [McGaugh(1991)]{McGaugh1991ApJ380}
          McGaugh, S. S. 1991, \apj, 380, 140

\bibitem [Miller \& Hodge(1996)]{MillerHodge1996ApJ458}
          Miller, B. W., \& Hodge, P. 1996, \apj, 458, 467                    

\bibitem [Moles et al.(1990)]{Molesetal1990AA228}   
          Moles, M., Aparicio, A., \& Masegosa, J. 1990, \aap, 228, 310      

\bibitem [Moustakas \& Kennicutt(2006a)]{MoustakasKennicutt2006ApJS164} 
          Moustakas, J., \& Kennicutt, R. C. 2006a, \apjs, 164, 81 

\bibitem [Moustakas \& Kennicutt(2006b)]{MoustakasKennicutt2006ApJ651} 
          Moustakas, J., \& Kennicutt, R. C. 2006b, \apj, 651, 155 

\bibitem [Moustakas et al.(2010)]{Moustakasetal2010ApJS190} 
          Moustakas, J., Kennicutt, R. C., Tremonti, C. A., et al. 2010, \apjs, 190, 233 

\bibitem [M\"{u}ller \& H\"{o}flich(1994)]{MullerHoflich1994AA281} 
          M\"{u}ller, E., \& H\"{o}flich, P. 1994, \aap, 281, 51 

\bibitem [Mundell et al.(2004)]{Mundelletal2004ApJ614}   
          Mundell, C. G., James, P. A., Loiseau, N., Schinnerer, E., \& Forbes, D. A.
          2004, \apj, 614, 648                                  

\bibitem [Mundell et al.(1995)]{Mundelletal1995MNRAS277} 
          Mundell, C. G., Pedlar, A., Axon, D. J., Meaburn, J., \& Unger, S. W. 
          1995, \mnras, 277, 641                                

\bibitem [Noeske et al.(2000)]{Noeskeetal2000AA361}      
          Noeske, K. G., Guseva, N. G., Fricke, K. J., et al.  2000, \aap, 361, 33                                  

\bibitem [Oey \& Kennicutt(1993)]{OeyKennicutt1993ApJ411}
          Oey, M. S., \& Kennicutt, R. C. 1993, \apj, 411, 137 

\bibitem [Olivares et al.(2010)]{Olivaresetal2010ApJ715} 
          Olivares, E., Hamuy, M., Pignata, G., et al. 2010, \apj, 715, 833 

\bibitem [Ondrechen \& van der Hulst(1989)]{OndrechenHulst1989ApJ342}
          Ondrechen, M. P., \& van der Hulst, J. M. 1989, \apj, 342, 29  

\bibitem [Ondrechen et al.(1989)]{Ondrechenetal1989ApJ342}
          Ondrechen, M. P., van der Hulst, J. M., \& Hummel, E. 1989, \apj, 342, 39  

\bibitem [Pagel(1997)]{pagel1997} 
          Pagel, B. E. J. 1997, Nucleosynthesis and Chemical Evolution of Galaxies
         (Cambridge: Cambridge Univ. Press)

\bibitem [Pagel et al.(1979)]{Pageletal1979MNRAS189} 
          Pagel, B. E. J., Edmunds, M. G., Blackwell, D. E., Chun, M. S., \& Smith, G. 1979, \mnras, 189, 95

\bibitem [Pagel et al.(1980)]{Pageletal1980MNRAS193} 
          Pagel, B. E. J., Edmunds, M. G., \& Smith, G. 1980, \mnras, 193, 219 

\bibitem [Pancoast et al.(2010)]{Pancoastetal2010ApJ723}  
          Pancoast, A., Sajina, A., Lacy, M., Noriega-Crespo, A., \& Rho, J. 
          2010, \apj, 723, 530   

\bibitem [Pastoriza et al.(1993)]{Pastorizaetal1993MNRAS260} 
          Pastoriza, M. G., Dottori, H. A., Terlevich, E., Terlevich, R., \& D\'{i}az, A. I.  
          1993, \mnras, 260, 177 

\bibitem [Patterson et al.(2012)]{Pattersonetal2012MNRAS422} 
          Patterson, M. T., Walterbos, R. A. M., Kennicutt, R. C., Chiappini, C., \& Thilker, D. A. 
          2012, \mnras, 422, 401 

\bibitem [Paturel et al.(2003)]{Patureletal2003AA412}  
          Paturel, G., Petit, C., Prugniel, P., et al.  2003, \aap, 412, 45 

\bibitem [Peletier et al.(1999)]{Peletieretal1999ApJS125}
          Peletier, R. F., Knapen, J. H., Shlosman, I., et al. 1999, \apjs, 125, 363

\bibitem [P\'{e}rez-Montero et al.(2009)]{PerezMonteroetal2009AA497}
          P\'{e}rez-Montero, E., Garc\'{i}a-Benito, R., D\'{i}az, A. I., P\'{e}rez, E., \& Kehrig, C. 
          2009, \aap, 497, 53

\bibitem [Petrosian et al.(2007)]{Petrosianetal2007ApJS170}
          Petrosian, A., McLean, B., Allen, R. J., \& MacKenty, J. W. 
          2007, \apjs, 170, 33 

\bibitem [Pettini \& Pagel(2004)]{PettiniPagel2004MNRAS348}
          Pettini, M., \& Pagel, B. E. J. 2004, \mnras, 348, L59

\bibitem [Phillips et al.(1984)]{Phillipsetal1984MNRAS210} 
          Phillips, M. M., Pagel, B. E. J., Edmunds, M. G., \& D\'{i}az, A. 1984, \mnras, 210, 701

\bibitem [Pietrzy\'{n}ski et al.(2010)]{Pietrzynskietal2010AJ140} 
          Pietrzy\'{n}ski, G., Gieren, W., Hamuy, M., et al. 2010, \aj, 140, 1475 

\bibitem [Pilyugin(1993)]{Pilyugin1993AA277}
          Pilyugin, L. S. 1993, \aap, 277, 42 

\bibitem [Pilyugin(2000)]{Pilyugin2000AA362}
          Pilyugin, L. S. 2000, \aap, 362, 325

\bibitem [Pilyugin(2001)]{Pilyugin2001AA369}
          Pilyugin, L. S. 2001, \aap, 369, 594

\bibitem [Pilyugin(2003)]{Pilyugin2003AA397}
          Pilyugin, L. S. 2003, \aap, 397, 109

\bibitem [Pilyugin et al.(2012)]{Pilyuginetal2012MNRAS424} 
          Pilyugin, L. S., Grebel, E. K., \& Mattsson, L. 2012, \mnras, 424, 2316

\bibitem [Pilyugin et al.(2014)]{Pilyugin2014AJ} 
          Pilyugin, L. S., Grebel, E. K., Zinchenko, A. I., \& Kniazev, A. Y. 2014, \aj, submitted 

\bibitem [Pilyugin et al.(2013)]{Pilyugin2013MNRAS432} 
          Pilyugin, L. S., Lara-L\'{o}pez, M.A., Grebel, E. K., et al., 2013, \mnras, 432, 1217

\bibitem [Pilyugin \& Mattsson(2011)]{PilyuginMattsson2011MNRAS412} 
          Pilyugin, L. S., \& Mattsson, L. 2011, \mnras, 412, 1145

\bibitem [Pilyugin \& Thuan(2005)]{PilyuginThuan2005ApJ631}
          Pilyugin, L. S., \& Thuan, T. X. 2005, \apj, 631, 231

\bibitem [Pilyugin \& Thuan(2011)]{PilyuginThuan2011ApJ726} 
          Pilyugin, L. S., \& Thuan, T. X. 2011, \apj, 726, L23 

\bibitem [Pilyugin et al.(2003)]{Pilyuginetal2003AA397} 
          Pilyugin, L. S., Thuan, T. X., \& V\'{\i}lchez, J. M. 2003, \aap, 397, 487

\bibitem [Pilyugin et al.(2006)]{Pilyuginetal2006MNRAS367} 
          Pilyugin, L. S., Thuan, T. X., \& V\'{\i}lchez, J. M. 2006, \mnras, 367, 1139 

\bibitem [Pilyugin et al.(2007)]{Pilyuginetal2007MNRAS376} 
          Pilyugin, L. S., Thuan, T. X., \& V\'{\i}lchez, J. M. 2007, \mnras, 376, 353 

\bibitem [Pilyugin et al.(2004)]{Pilyuginetal2004AA425} 
          Pilyugin, L. S., V\'{\i}lchez, J. M., \& Contini, T. 2004, \aap, 425, 849 

\bibitem [Pilyugin et al.(2010)]{Pilyuginetal2010ApJ720}
          Pilyugin, L. S., V\'{\i}lchez, J. M., \& Thuan, T. X. 2010, \apj, 720, 1738 

\bibitem [Pisano et al.(1998)]{Pisanoetal1998AJ115} 
          Pisano, D. J., Wilcots, E. M., \& Elmegreen, B. G. 1998, \aj, 115, 975 

\bibitem [Pohlen \& Trujillo(2006)]{PohlenTrujillo2006AA454} 
          Pohlen, M., \& Trujillo, I. 2006, \aap, 454, 759 

\bibitem [Poznanski et al.(2009)]{Poznanskietal2009ApJ694} 
          Poznanski, D., Butler, N., Filippenko, A. V., et al. 2009, \apj, 694, 1067 

\bibitem [Puche et al.(1991)]{Pucheetal1991AJ101} 
          Puche, D., Carignan, C., \& Wainscoat, R. J. 1991, \aj, 101, 447 

\bibitem [Puche et al.(1992)]{Pucheetal1992AJ103} 
          Puche, D., Westpfahl, D., \& Brinks, E. 1992, \aj, 103, 1841    

\bibitem [Rayo et al.(1982)]{Rayoetal1982ApJ255} 
          Rayo, J.F., Peimbert, M., \& Torres-Peimbert, S. 1982, \apj, 255, 1

\bibitem [Rela\~{n}o et al.(2010)]{Relanoetal2010MNRAS402} 
          Rela\~{n}o, M., Monreal-Ibero, A., V\'{i}lchez, J. M., \& Kennicutt, R. C. 
          2010, \mnras, 402, 1635 

\bibitem [Riad et al.(2010)]{Riadetal2010MNRAS401} 
          Riad, I. F., Kraan-Kortweg, R. C., \& Woudt, P. A. 2010, \mnras, 401, 924  

\bibitem [Richer et al.(2001)]{Richeretal2001AA370} 
          Richer, M. G., Bullejos, A., Borissova, J., et al. 2001, \aap, 370, 34 

\bibitem [Robertson et al.(2012)]{Robertsonetal2012ApJ748}	
          Robertson, P., Shields, G. A., \& Blanc, G. A. 2012, \apj, 748, 48  

\bibitem [Rodrigues et al.(1998)]{Rodriguesetal1998AAS128}	
          Rodrigues, I., Dottori, H., Cepa, J., \& V\'{i}lchez, J. M. 1998, \aaps, 128, 545

\bibitem [Rosales-Ortega et al.(2010)]{RosalesOrtegaetal2010MNRAS405}	
          Rosales-Ortega, F. F., Kennicutt, R. C., S\'{a}nchez, S. F.,  et al. 2010, \mnras, 405, 735    

\bibitem [Rosales-Ortega et al.(2011)]{RosalesOrtegaetal2011MNRAS415}	
          Rosales-Ortega, F. F.,  D\'{i}az, A. I., Kennicutt, R. C., \& S\'{a}nchez, S. F. 
          2011, \mnras, 415, 2439  

\bibitem [Rownd et al.(1994)]{Rowndetal1994AJ108}	
          Rownd, B. K., Dickey, J. M., \& Helou, G. 1994, \aj, 108, 1638 

\bibitem [Roy \& Walsh(1997)]{RoyWalsh1997MNRAS288} 
          Roy, J.-R., \& Walsh, J. R.  1997, \mnras, 288, 715 

\bibitem [Roy et al.(1991)]{Royetal1991AJ101}     
          Roy, J.-R., Wang, J., \& Arsenault, R. 1991, \aj, 101, 825 

\bibitem [Rubin et al.(1988)]{Rubinetal1988ApJ333}	
          Rubin, V. C., Whitmore, B. C., \& Ford, W. K. 1988, \apj, 333, 522 

\bibitem [Ryder(1995)]{Ryder1995ApJ444}	
          Ryder, S. D. 1995, \apj, 444, 610  

\bibitem [Saha et al.(2002)]{Sahaetal2002AJ124}	
          Saha, A., Claver, J., \& Hoessel, J. G. 2002, \aj, 124, 839  

\bibitem [Saha et al.(2006)]{Sahaetal2006ApJS165}	
          Saha, A., Thim, F., Tamman, G. A., Reindl, B., \& Sandage, A. 2006, \apjs, 165, 108  

\bibitem [Sakai et al.(1999)]{Sakaietal1999ApJ511}  
          Sakai, S., Madore, B. F., \& Freedman, W. L. 1999, \apj, 511, 671 

\bibitem [S\'{a}nchez et al.(2012)]{Sanchezetal2012AAa000} 
          S\'{a}nchez, S.F., Rosales-Ortega, F. F., Marino, R.A., et al. 2012, \aap, 546, A2

\bibitem [Sandage(1986)]{Sandage1986AA161} 
          Sandage, A. 1986, \aap, 161, 89  

\bibitem [Sanders et al.(2012)]{Sandersetal2012ApJ758}  
          Sanders, N. E., Caldwell, N., McDowell, J., \& Harding, P. 2012, \apj, 758, 133 

\bibitem [Schmidt et al.(1994)]{Schmidtetal1994ApJ432} 
          Schmidt, B. P., Kirshner, R. P., Eastman, R. G., et al. 1994, \apj, 432, 42 

\bibitem [Searle(1971)] {Searle1971ApJ168} 
          Searle, L. 1971, \apj, 168, 327 

\bibitem [Sedwick \& Aller(1981)] {SedwickAller1981PNAS78}
          Sedwick, K. E., \& Aller, L. H. 1981, Proc. Nat. Acad. Sci. USA., 78, 1994

\bibitem [Send(1982)] {Send1982AA112}     
          Send, U. 1982, \aap, 112, 235 

\bibitem [Seth et al.(2005)] {Sethetal2005AJ129}     
          Seth, A. C., Dalcanton, J. J., \& de Jong, R. S. 2005, \aj, 129, 1331        

\bibitem [Sharina et al.(1997)] {Sharinaetal1997AstronLett23}
          Sharina, M. E., Karachentsev, I. D., \& Tikhonov, N. A. 1997, Astron. Lett., 23, 373

\bibitem [Shields et al.(1991)] {Shieldsetal1991ApJ371} 
          Shields, G. A., Skillman, E. D., \& Kennicutt, R. C. 1991, \apj, 371, 82 

\bibitem [Skillman(1985)]{Skillman1985ApJ290} 
          Skillman, E. D. 1985, \apj, 290, 449 

\bibitem [Skillman et al.(1996)]{Skillmanetal1996ApJ462}  
          Skillman, E. D., Kennicutt, R. C., Shields, G. A., \& Zaritsky, D. 1996, \apj, 462, 147 

\bibitem [Skillman et al.(2013)]{Skillmanetal2013ApJ000}  
          Skillman E. D., Salzer, J. J., Berg, D. A., et al. 2013, \aj, 146, 3 

\bibitem [Smith(1975)]{Smith1975ApJ199} 
          Smith, H. E. 1975, \apj, 199, 591 

\bibitem [Staghellini et al.(2010)]{Stanghellinietal2010AA521} 
          Stanghellini, L., Magrini, L., Villaver, E., \& Galli, D. 2010, \aap, 521, A3 

\bibitem [Stasi\'{n}ska (2006)]{Stasinska2006AA454}
          Stasi\'{n}ska, G.  2006, \apj, 454, L127 

\bibitem [Stasi\'{n}ska et al.(1986)]{Stasinskaetal1986AA154} 
          Stasi\'{n}ska, G., Comte, G., \& Vigroux, L. 1986, \aap, 154, 352

\bibitem [Storchi-Bergmann et al.(1996a)]{StorchiBergmannetal1996ApJ472}  
          Storchi-Bergmann, T., Rodr\'{i}guez-Ardila, A., Schmidt, H. R., Wilson, A. S., \& Baldwin, J. A. 
          1996a, \apj, 472, 83 

\bibitem [Storchi-Bergmann et al.(1996b)]{StorchiBergmannetal1996ApJ460}  
          Storchi-Bergmann, T., Wilson, A. S., \& Baldwin, J. A. 1996b, \apj, 460, 252

\bibitem [Storey \&  Zeippen(2000)]{StoreyZeipen2000MNRAS312}
          Storey, P. J., \& Zeippen, C. J. 2000, \mnras, 312, 813

\bibitem [Springob et al.(2009)]{Springobetal2009ApJS182} 
          Springob, C. M., Masters, K. L., Haynes, M. P., Giovanelli, R., \& Marinoni, C. 
          2009, \apjs, 182, 474  

\bibitem [Stauffer \& Bothun(1984)]{StaufferBothun1984AJ89} 
          Stauffer, J. R., \& Bothun, G. D. 1984, \aj, 89, 1702 

\bibitem [Swaters \& Balcells(2002)]{SwatersBalcells2002AA390}  
          Swaters, R. A., \& Balcells, M. 2002, \aap, 390, 863  

\bibitem [Terry et al.(2002)]{Terryetal2002AA393}   
          Terry, J. N., Paturel, G., \& Ekholm, T. 2002, \aap, 393, 57 

\bibitem [Thim et al.(2004)]{Thimetal2004AJ127}  
          Thim, F., Hoessel, J. G., Saha, A., et al. 2004, \aj, 127, 2322

\bibitem [Thuan et al.(2010)]{Thuanetal2010ApJ712}
          Thuan, T. X., Pilyugin, L. S., \& Zinchenko, I. A. 2010, \apj, 712, 1029  

\bibitem [Tikhonov \& Galazoutdinova(2002)]{TikhonovGalazoutdinova2002Afz45}  
          Tikhonov, N. A., \& Galazoutdinova, O. A. 2002, Afz, 45, 253 

\bibitem [Torres-Peimbert et al.(1989)]{TorresPeimbert1989ApJ345}
          Torres-Peimbert, S., Peimbert, M., \& Fierro, J. 1989, \apj, 345, 186

\bibitem [Tremonti et al.(2004)]{Tremonti2004ApJ613} 
          Tremonti, C. A., Heckman, T.M., Kauffmann, G., et al.,  2004, \apj, 613, 898


\bibitem [T\"{u}llmann et al.(2003)]{Tullmannetal2003AA412}	
          T\"{u}llmann, R., Rosa, M. R., Elwert, T., et al. 2003, \aap, 412, 69 

\bibitem [Tully(1988)]{Tully1988catalogue}	
          Tully, R. B. 1988, Nearby Galaxy Catalogue (Cambridge Univ. Press, Cambridge)

\bibitem [Tully et al.(1996)]{Tullyetal1996AJ112} 
          Tully, R. B., Verheijen, M. A. W., Pierce, M. J., Huang, J.-S., \& Wainscoat, R. J. 
          1996, \aj, 112, 2471 

\bibitem [Tully et al.(2009)]{Tullyetal2009AJ138}	
          Tully, R. B., Rizzi, L., Shaya, E. J., et al.  2009, \aj, 138, 323  

\bibitem [van Albada(1980)]{vanAlbada1980AA90} 
          van Albada, G. D. 1980, \aap, 90, 123  

\bibitem [van den Bergh(2008)]{vandenBergh2008AA490}  
          van den Bergh, S. 2008, \aap, 490, 97 

\bibitem [van den Hoek \& Groenewegen(1997)]{vandenhoek1997} 
          van den Hoek, L. B., \& Groenewegen, M. A. T. 1997, \aaps, 123, 305

\bibitem [van Zee(2000)]{vanZee2000AJ119}     
          van Zee, L. 2000, \aj, 119, 2757

\bibitem [van Zee \& Bryant(1999)]{vanZeeBryant1999AJ118} 
          van Zee, L., \& Bryant, J. 1999, \aj, 118, 2172  

\bibitem [van Zee  \& Haynes(2006)]{vanZeeHaynes2006ApJ636}
          van Zee, L., \& Haynes, M. P. 2006, \apj, 636, 214 

\bibitem [van Zee et al.(1998)]{vanZeeetal1998AJ116} 
          van Zee, L., Salzer, J. J., Haynes, M. P., O`Donoghue, A. A., \& 
          Balonek, T. J.  1998, \aj, 116, 2805

\bibitem [van Zee  et al.(2006)]{vanZeeetal2006ApJ637}  
          van Zee, L., Skillman, E. D., \& Haynes, M. P. 2006, \apj, 637, 269 

\bibitem [Verdes-Montenegro et al.(2000)]{VerdesMontenegro2000AA356}  
          Verdes-Montenegro, L., Bosma, A., \& Athanassoula, E. 2000, \aap, 356, 827

\bibitem [Vila-Costas \& Edmunds(1992)]{VilaCostas1992MNRAS259} 
          Vila-Costas, M. B., \& Edmunds, M. G. 1992, \mnras, 259, 121

\bibitem [Vilardell et al.(2010)]{Vilardellietal2010AA509}
          Vilardell, F., Ribas, I., Jordi, C., Fitzpatrick, E. L., \& Guinan, E. F. 
          2010, \aap, 509, A70

\bibitem [V\'{i}lchez et al.(1988a)]{Vilchezetal1988PASP100}
          V\'{i}lchez, J. M., Edmunds, M. G., \& Pagel, B. E. J.  1988b, \pasp, 100, 1428 

\bibitem [V\'{i}lchez et al.(1988b)]{Vilchezetal1988MNRAS235}
          V\'{i}lchez, J. M., Pagel, B. E. J., D\'{i}az, A. I., Terlevich, E., \& Edmunds, M. G. 
          1988a, \mnras, 235, 633  

\bibitem [Walsh \& Roy(1997)]{WalshRoy1997MNRAS288} 
          Walsh, J. R., \& Roy, J.-R. 1997, \mnras, 288, 726 

\bibitem [Walter et al.(2008)]{Walteretal2008AJ136} 
          Walter, F., Brinks, E., de Blok, W. J. G., et al. 2008, \aj, 136, 2563   

\bibitem [Walterbos \& Kennicutt(1987)]{WalterbosKennicutt1987AAS69} 
          Walterbos, R. A. M., \& Kennicutt, R. C. 1987, \aaps, 69, 311

\bibitem [Webster \& Smith(1983)]{WebsterSmith1983MNRAS204} 
          Webster, B. L., \& Smith, M. G. 1983, \mnras, 204, 743

\bibitem [Werk et al.(2011)]{Werketal2011ApJ735} 
          Werk, J. K., Putman, M. E., Meurer, G. R., \& Santiago-Figueroa, N. 
          2011, \apj, 735, 71

\bibitem [Whiteoak \& Gardner(1977)]{WhiteoakGardner1977AJP30} 
          Whiteoak, J. B., \& Gardner, F. F. 1977, Aust. J. Phys., 30, 187 

\bibitem [Wood-Vasey et al.(2008)]{WoodVaseyetal2008ApJ689} 
          Wood-Vasey, W. M., Friedman, A. S., Bloom, J. S., et al. 2008, \apj, 689, 377 

\bibitem [York et al.(2000)]{yorketal2000AJ120} 
          York, D. G., Adelman, J., Anderson, J. E., et al.  2000, \aj, 120, 1579 

\bibitem [Zahid \& Bresolin(2011)]{ZahidBresolin2011ApJ141}
          Zahid, H. J., \& Bresolin, F. 2011, \aj, 141, 192

\bibitem [Zaritsky et al.(1989)]{Zaritskyetal1989AJ97}
          Zaritsky, D., Elston, R., \& Hill, J. M. 1989, \aj, 97, 97  

\bibitem [Zaritsky et al.(1990)]{Zaritskyetal1990AJ99}
          Zaritsky, D., Elston, R., \& Hill, J. M. 1990, \aj, 99, 1108

\bibitem [Zaritsky et al.(1994)]{Zaritskyetal1994ApJ420} 
          Zaritsky, D., Kennicutt, R. C., \& Huchra, J. P. 1994, \apj, 420, 87 

\bibitem [Zurita \& Bresolin(2012)]{ZuritaBresolin2012MNRAS427}
          Zurita, A., \& Bresolin, F. 2012, \mnras, 427, 1463 

\end{thebibliography}
\end{document}